\documentclass[aps,prx,10pt,twocolumn,superscriptaddress,amsmath,amssymb,longbibliography,floatfix]{revtex4-2}
\usepackage[english]{babel}
\usepackage{graphicx,bbm}
\usepackage[utf8]{inputenc}
\usepackage{soul}
\usepackage[colorlinks,linkcolor=blue,citecolor=blue,urlcolor=blue]{hyperref}
\usepackage{physics}
\usepackage{dsfont}

\begin{document} 
\title{Graph-Theoretic Detection of Hilbert Space Fragmentation}
\author{A. Rutkowski}
\author{M. Mierzejewski}
\author{J. Herbrych}
\affiliation{Institute of Theoretical Physics, Faculty of Fundamental Problems of Technology, Wroc{\l}aw University of Science and Technology, 50-370 Wroc{\l}aw, Poland}
\date{\today}

\begin{abstract}
Hilbert-space fragmentation provides a mechanism for ergodicity breaking in quantum many-body systems even in the absence of disorder, leading to dynamically disconnected sectors and strong memory of initial conditions. However, identifying such structures is often challenging and typically relies on prior knowledge of conservation laws or model-specific analytical insight. Here we introduce an unbiased approach based on spectral graph theory and, within this framework, formulate the concept of nearly fragmented systems, in which perturbative processes couple otherwise fragmented sectors while preserving their dynamical imprint. By representing basis states as vertices and Hamiltonian matrix elements as edges, we map the connectivity structure of the many-body Hilbert space onto a graph and analyze it using tools such as the Laplacian spectrum, Fiedler vectors, and modularity. Exact fragmentation corresponds to disconnected graph components, while nearly fragmented systems manifest as weakly connected communities whose structure can still be resolved spectrally. Applying this framework to the one-dimensional $t$-$J$ model and its perturbations, we demonstrate that graph-theoretic diagnostics reliably identify both fragmented and nearly fragmented Hilbert-space structures and capture the hierarchy of dynamical time scales that governs the system’s evolution. We further show that the method extends beyond kinetically constrained models by applying it directly to the Hubbard chain, where it reveals the emergence of nearly decoupled subspaces associated with doublon dynamics and spin configurations. Our results establish the spectral graph analysis as a general and scalable tool for diagnosing fragmentation and approximate dynamical constraints in complex quantum many-body systems.
\end{abstract}
\maketitle

\section{Introduction}
Understanding how and when isolated quantum many-body systems thermalize is a central problem of nonequilibrium physics. In generic interacting systems, ergodicity implies that the long-time dynamics explores the entire Hilbert space consistent with global conservation laws, leading to thermal expectation values described by statistical ensembles. This paradigm is formalized through the eigenstate thermalization hypothesis (ETH) \cite{Deutsch1991,Srednicki1994,Rigol2008,Alessio2016}, which posits that individual many-body eigenstates already encode thermal properties. However, a growing body of work has revealed robust mechanisms by which ergodicity can fail even in the absence of quenched disorder.

One such mechanism is Hilbert space fragmentation, in which the many-body Hilbert space decomposes into dynamically disconnected sectors that are not related by conventional symmetries. Unlike integrable systems, where extensive local conserved quantities constrain the dynamics while still allowing exploration within a large manifold, fragmented systems split into exponentially many subspaces that are mutually inaccessible under the Hamiltonian action. As a consequence, dynamics initialized in one sector remain confined to it, leading to strong memory effects and nonthermal steady states. One may also distinguish between weak and strong fragmentation \cite{Sala2020,Khemani2020}, corresponding to cases where typical initial states do and do not thermalize, respectively. Such classification is partially motivated by weak and strong versions of the ETH \cite{Alessio2016,Beugeling2014}. These two types of fragmentation correspond to how the dimension of the largest subspaces compares with that of the full Hilbert space. In the case of strong fragmentation, all subspaces are exponentially smaller than the Hilbert space.

Fragmentation can originate from kinematic constraints or conservation laws that act locally rather than globally. Prominent examples include kinetically constrained models~\cite{Yang2020,Khemani2020,Langlett2021,Pozsgay2021,Brighi2023,Vanja2026,Nicolau2026}, dipole-conserving systems~\cite{Herviou2021,Nandy2024}, certain limits of lattice fermion models \cite{Tomasi2019,Scherg2021,Nandy2024}, systems with long-range interactions \cite{Yang2025,Prodius2026}, and even open systems~\cite{Li2023}. In all of these cases, the fragmentation is exact: the Hamiltonian matrix acquires a block-diagonal structure in a suitable (computational) basis. It should, however, be noted that identifying fragmented subspaces -- or even establishing their existence -- is in general a nontrivial task. For this reason, it is particularly worth emphasizing experimental attempts to identify nearly fragmented systems in cold atom experiments \cite{Scherg2021,Kohlert2023,Adler2024,Honda2025,Zhao2025,Karch2025} and quantum emulators \cite{Guo2021}. 

The impact of integrability on the dynamics of physical observables is most thoroughly described by local integrals of motion. A similar, though not identical, concept can be applied to systems with fragmentation by introducing the so-called statistically localized integrals of motion (SLIOMs) \cite{Rakovszky2020,Lydzba2024}. It was shown \cite{Moudgalya2021,Moudgalya2022-1} that commutant algebras -- defined as the algebra of operators commuting with all local generators of the dynamics of the system -- can be used to identify SLIOMs. However, this approach typically requires cumbersome calculations. Alternatively, one may numerically test whether a given Hilbert space decomposes into dynamically disconnected Krylov subspaces \cite{Moudgalya2022-2} by iteratively applying the Hamiltonian, $H$, to a basis state, $\ket{b}$, to construct a Krylov subspace ${\cal H}_{\alpha}={\rm span}(\ket{b},H\ket{b}, H^2\ket{b},\dots )$, and repeating the procedure for other basis states, $\ket{b'} \notin {\cal H}_{\alpha}$, until the full Hilbert space ${\cal H}=\bigoplus_{\alpha=1}^F {\cal H}_{\alpha}$ is reconstructed (with $F$ as a number of subspaces). The resulting ${\cal H}_{\alpha}$ are dynamically disconnected in that $H\ket{\psi} \in {\cal H}_{\alpha}$ for any $\ket{\psi} \in {\cal H}_{\alpha}$. Equivalent observation is that the matrix elements vanish, $\bra{b} H\ket{b'}=0$, whenever $\ket{b}$ and $\ket{b'}$ belong to different subspaces, ${\cal H}_{\alpha}\ne {\cal H}_{\alpha'}$

Given a Hamiltonian, without assuming any prior knowledge about its structure or conserved quantities, one may ask whether and how the Hilbert space can be divided into disconnected subspaces such that the dynamics within each subspace is fast, whereas the dynamics between different subspaces is slow or frozen. In this work, we propose an unbiased approach to such a problem based on spectral graph theory. Graph representations provide a versatile framework for analyzing complex systems across many areas of science. In such approaches, the elements of a system are represented as vertices, while their interactions or relations define edges of a graph. This abstraction allows a wide variety of problems to be studied using a common mathematical language \cite{Boccaletti2006,Fortunato2010}. For instance, graph-theoretic methods are widely used to analyze large-scale infrastructure and communication networks \cite{Newman2006-1,Newman2006-2}, characterize community structure in social networks, and investigate connectivity patterns in biological systems, such as gene regulatory and neural networks \cite{Jeong2000,Alon2007}. In physics and chemistry, graphs naturally arise in the study of molecular structures, transport processes on networks, and percolation phenomena \cite{Watts1998,Albert2002}, as well as in the study of slow dynamics in systems with kinetic constraints \cite{Menzler2025,Menzler2026}.

Our key observation is that spectral properties of graph matrices have proven particularly useful for detecting connectivity patterns, clustering, and collective behavior in complex systems \cite{Chung1997,Spielman2007}. Because these tools provide direct information about the connectivity structure of a system, they offer a natural and unbiased way to analyze configuration spaces generated by many-body Hamiltonians and to identify disconnected or weakly connected sectors, i.e., fragmentation-like phenomena in the Hilbert space. In such language, the connectivity structure induced by the Hamiltonian in a chosen basis naturally defines a graph: basis states correspond to vertices, and nonzero Hamiltonian matrix elements define edges. Exact Hilbert space fragmentation then corresponds to the graph decomposing into disconnected components (often called communities). In 1973, Czech mathematician Miroslav Fiedler \cite{Fiedler1973} proposed a simple algorithm for identifying such disconnected subspaces. Consequently, this reformulation of the problem allows us to leverage mature graph-theory tools to diagnose fragmentation. 

Furthermore, we show that a useful analogy can be drawn with the well-known distinction between integrable and nonintegrable systems. Integrable models possess an extensive set of conserved quantities that strongly constrain the dynamics. When integrability is weakly broken, these constraints cease to be exact, yet their imprint often persists as approximate conservation laws and long-lived dynamical structures,  known as prethermalization \cite{Kollar2011,Bertini2015,Mallayya2019}. We will argue that an analogous hierarchy emerges in the context of Hilbert-space fragmentation. In fully fragmented systems, the Hilbert space decomposes into many dynamically disconnected sectors, so that states belonging to different sectors cannot be connected by the Hamiltonian dynamics. As a result, the system retains a strong memory of the initial conditions and fails to explore the full Hilbert space. In contrast, nearly fragmented systems arise when weak processes couple these sectors, formally restoring connectivity of the Hilbert space. Despite this, the underlying structure of weakly connected subspaces may still organize the dynamics and leave signatures in the system's behavior.

By combining physical insight into constrained dynamics with spectral graph-theoretic tools, we establish a systematic framework to detect both exact and approximate Hilbert space fragmentation in quantum many-body systems. We demonstrate its effectiveness for the Hubbard and $t$-$J$ models and their perturbations, and argue that the method generalizes to a broad class of constrained quantum many-body systems beyond kinetically induced fragmentation. The paper is structured as follows: In Sec.~\ref{sec:model}, we introduce the one-dimensional $t$-$J_{z}$ model and its basic properties from the perspective of the Hilbert space fragmentation. Sec.~\ref{sec:gt} is devoted to the main tool of our investigation: the spectral analysis of the graphs. Especially, in Sec.~\ref{sec:fiedler} and Sec.~\ref{sec:modularity}, we introduce the Fiedler vectors and graph modularity, respectively. In Sec.~\ref{sec:results}, we present an analysis of the fragmented and nearly-fragmented $t$-$J$-like models, while in Sec.~\ref{sec:dynamics}, we discuss how the latter influences the system's dynamics. Finally, in Sec.~\ref{sec:hubbard}, we discuss how graph-theoretic tools can be used to analyze the Hubbard model. Discussion and Conclusions are given in Sec.~\ref{sec:conclusions}.

\section{Model}
\label{sec:model}
We base our investigation of Hilbert space fragmentation on the one-dimensional $t$-$J$ model. However, as we will show below, our conclusions extend more broadly to systems in which kinematic constraints generate multiple dynamical time scales. The $t$-$J$ model is defined as 
\begin{eqnarray}
H&=& H_t + H_J\,,\\
H_t&=&t \sum_{\sigma,l}\left(P_{G}\,c^\dagger_{l,\sigma}c^{\phantom{\dagger}}_{l+1,\sigma}P_G+\mathrm{H.c.}\right)\,,\nonumber\\
H_J&=&J\sum_l\left[\frac{1}{2}\left(S^+_{l}S^-_{l+1}+\mathrm{H.c.}\right)+S^z_{l}S^z_{l+1}\right]\,.\nonumber
\label{eq:tj}
\end{eqnarray}
Here, $c^\dagger$ ($c$) represent fermionic creation (annihilation) operators, while $S^+$/$S^-$ and $S^zS^z$ denote spin raising/lowering operators and nearest-neighbor spin-spin interactions, respectively. The $t$-$J$ model is typically derived via second-order perturbation theory in the large-interaction (Mott-insulating) limit of the doped Hubbard model. Consequently, double occupancy is forbidden, and the on-site Hilbert space is restricted to holes $\ket{0}$ and singly occupied states with spin projections $\ket{\uparrow}$ and $\ket{\downarrow}$. The latter is achieved via the Gutzwiller projector $P_G=\prod_\ell (1-n_{\ell\uparrow}n_{\ell\downarrow})$ with $n_{\ell\sigma}=c^\dagger_{\ell,\sigma}c^{\phantom{\dagger}}_{\ell,\sigma}$. In the following, we consider systems of $L$ sites near half-filling, set $t=1$ as energy unit, and denote the number of holes by $N_h$. Furthermore, we will consider open boundary conditions and sectors with vanishing total magnetization $\sum_\ell S^z_\ell=0$.

\begin{figure*}[!t]
\includegraphics[width=1.0\textwidth]{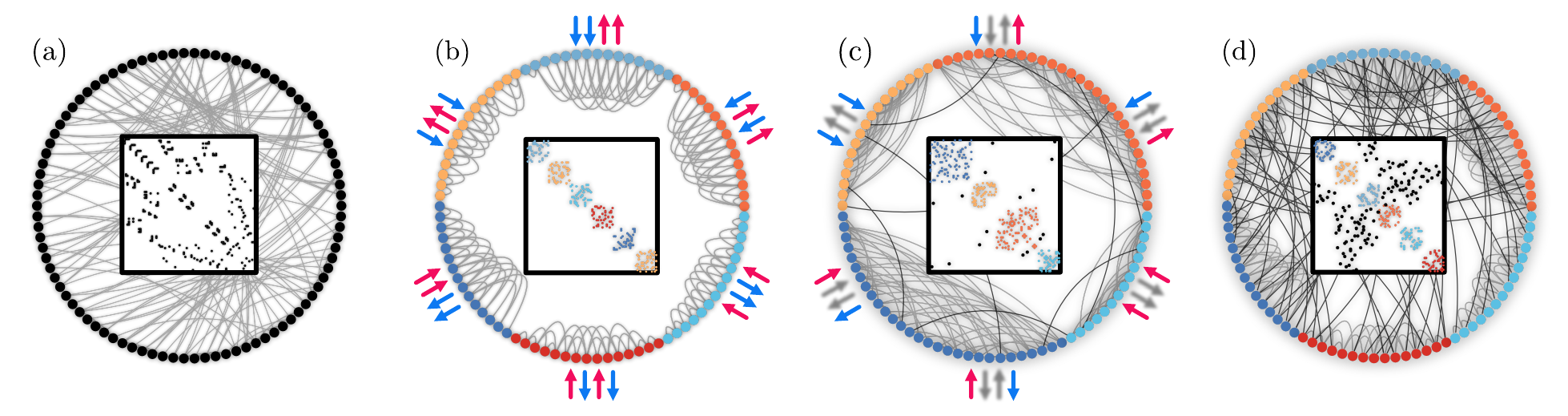}
\caption{Circular graph representation of Hamiltonian matrices of (a) fragmented $t$-$J_{z}$ model for $L=6$ sites, $N_h=2$ holes, and Hilbert space dimension $\dim({\cal H})=90$. Such a system (in $S^z_\mathrm{tot}=0$ magnetization sector) exhibits fragmentation into $6$ subspaces. Grey lines depict nonzero matrix elements (connections). (b) The same as (a) but with properly ordered basis states that reveal the Hilbert space fragmentation. (c) Graph of nearly fragmented rung-depleted $t$-$J_{z}$ ladder ($L=6$, $N_h=2$). $4$ subspaces detected by Fiedler vector analysis are marked with colors. In (b) and (c), spin configurations in squeezed space are also presented. (d) Graph of nearly fragmented $t$-$J$ model ($L=6$, $N_h=2$, and $t\gg J$). In (c,d), connections between subspaces are marked with black color. The insets represent nonzero elements of the corresponding Hamiltonian matrix.}
\label{fig1}
\end{figure*}

For $H_J=0$, the so-called $t$-$J_{z}$ model is realized, in which mobile holes cannot alter the overall spin configuration of the electrons, e.g., \mbox{$\ket{\uparrow0\uparrow\downarrow\downarrow0\uparrow}\rightarrow\ket{\uparrow\uparrow0\downarrow0\downarrow\uparrow}$}. Consequently, the spin configurations are frozen in the so-called squeezed space -- defined as the configuration of spins in real space with the holes removed (e.g., $\ket{\uparrow\uparrow\downarrow\downarrow\uparrow}$ in the example above) -- leading to a fragmentation of the Hilbert space into $2^{L-N_h}$ disconnected subspaces. All subspaces are uniquely labeled by the spin configurations in the squeezed space, i.e., by the spin projection of the first, second, and subsequent fermions. The latter can then hop over distances of up to $N_h$ sites. Note that such spin projections do not represent local conserved quantities, but rather SLIOMs \cite{Rakovszky2020,Moudgalya2022-1}. 

We want to point out that the $t$-$J$ model is typically defined with additional $-J/4\,n_{\ell}n_{\ell+1}$ term (with $n_\ell=n_{\ell\uparrow}+n_{\ell\downarrow}$), while the $t$-$J_{z}$ model has finite $S^z_{\ell}S^z_{\ell+1}$ interaction term. The model without the latter term is also called the $t$-model. Since such terms are diagonal in the chosen basis, neither is relevant to the system’s fragmentation, and we therefore omit them.

\section{Graph theory tools}
\label{sec:gt}

\subsection{Fiedler vector analysis}
\label{sec:fiedler}
Consider first a simple setup with two disjoint subspaces, ${\cal H}={\cal H}_{1}\oplus {\cal H}_{2}$. We assume that each basis state $\ket{b_i} \in {\cal H}$ belongs to one subspace and stays in this subspace during evolution, i.e., either $\exp(-iHt)\ket{b_i} \in {\cal H}_{1}$ or $\exp(-iHt)\ket{b_i} \in {\cal H}_{2}$. In graph theory language, such sets of basis states, $\{ \ket{b_i} \in {\cal H}_{1} \}$ and $\{ \ket{b_i} \in {\cal H}_{2} \}$ represent communities that are not connected by any edge of the graph. For the $t$-$J_{z}$ model described above, such a partition corresponds to two different spin configurations in squeezed space. Next, consider the adjacency matrix $A$ connecting the vertices $\ket{b_i}$ and $\ket{b_j}$, defined such that $A_{ij} = 1$ if $\bra{b_i}H\ket{b_j} \ne 0$, and $A_{ij} = 0$ otherwise. In graph-theory language, this means that $A_{ij}=1$ if there exists an edge between the vertices $\ket{b_i}$ and $\ket{b_j}$. For the $t$-$J_{z}$ model, the adjacency matrix corresponds to the kinetic part of the Hamiltonian, $A_{ij}=|\bra{b_i}H\ket{b_j}|^2$, which connects basis states that differ by a hole hopping process while preserving the spin configuration (note that here we have chosen $t=1$). Choosing the adjacency matrix as the square of the Hamiltonian matrix elements may appear somewhat arbitrary. In Sec.~\ref{sec:dynamics}, however, we show that this choice is well motivated by the actual quantum dynamics of systems with dynamically disconnected or nearly disconnected subspaces.

Since the ordering of the computational basis states $\ket{b_i}$ is typically arbitrary with respect to the underlying structure of the disjoint subspaces, the problem at hand reduces to finding an ordering of basis states that reveals the fragmentation (i.e., identifies which basis states belong to which subspace ${\cal H}_{1}$ or ${\cal H}_{2}$). See Fig.~\ref{fig1}(a,b). Fiedler showed \cite{Fiedler1973,NgJordanWeiss2002} that analyzing the eigenvalues of the Laplacian matrix $L=D-A$, where $D=\mathrm{diag}(d_1,d_2,\dots)$ with $d_{i}=\sum_j A_{ij}$ is a diagonal matrix representing the degree of each vertex, can reveal the underlying structure of the graph. Since the Laplacian matrix is symmetric and satisfies $\sum_j L_{ij}=0$, the eigenvalues of $L$,
\begin{equation}
\frac{\mathbf{v}^T L \mathbf{v}}{\mathbf{v}^T \mathbf{v}} = \lambda, 
\label{eq:laplacian}
\end{equation}
are positive semidefinite, $\lambda_i\geqslant 0$. Equation~\eqref{eq:laplacian} can be recast as
\begin{equation}
\mathbf{v}^T L \mathbf{v} = \frac{1}{2} \sum_{i,j} A_{ij} (v_i - v_j)^2\geqslant 0\,,
\label{fiedler}
\end{equation}
where $\mathbf{v}^T=(v_1,v_2,v_3,\dots)$ is a vector defined on the basis states (vertices) $\ket{b_i}=\ket{b_1}\,,\ket{b_2}\,,\ket{b_3}\,,\dots$. The above equation has at least one trivial solution (irrespective of the structure of the graph) with eigenvalue $\lambda_1=0$, corresponding to $\mathbf{v}_1^T=(1,1,1,\dots)$ (in the following, to visualize the problem better, we will consider orthogonal but non-normalized eigenvectors). For the case described above -- namely, two disjoint subspaces ${\cal H}_{1}$ and ${\cal H}_{2}$ -- there is another solution with $\lambda_2=0$ for a vector $\mathbf{v}_2^T=(v_1,v_2,v_3,\dots)$, where $v_i=1$ if $\ket{b_i}\in {\cal H}_{1}$ and $v_i=-1$ if $\ket{b_i}\in {\cal H}_{2}$. Indeed, since $A_{ij}=0$ for states belonging to different subspaces and $v_i - v_j=0$ for states inside the same subspace, the second eigenvector of $L$ (the so-called {\it Fiedler vector}) encodes the nontrivial structure of the graph. The above procedure can be generalized to any number $F$ of subspaces, ${\cal H}=\bigoplus_{\alpha=1}^F {\cal H}_{\alpha}$, yielding that the number of disjoint subspaces equals the degeneracy of the lowest eigenvalue of the Laplacian matrix, $F=\dim(\lambda_i=0)$. We refer the interested reader to the Appendix~\ref{sec:ap1} for more technical details about Fiedler vector analysis.

Note that the above conclusions are reached without any {\it a priori} knowledge of the fragmentation mechanism, other than the kinetic term of the Hamiltonian. Furthermore, this simple analysis yields few nontrivial insights: (i) if the number of zero eigenvalues is larger than one, $\dim(\lambda_i=0)>1$, the Hamiltonian is block-diagonal in the basis $\ket{b_i}$; (iii) the $F=\dim(\lambda_i=0)$ yields exact number of disconnected subspaces; and (iii) the analysis of Fiedler vectors $\mathbf{v}_i$ (with $\lambda_i=0$) reveals the block structure (fragmentation) of the system and provides a natural starting point for the construction of SLIOMs.

Interestingly, the analysis of the Laplacian can also be extended to graphs (systems) without strictly separated subspaces, i.e., to nearly fragmented systems \cite{Lisiecki2025}. In such cases, the condition for vanishing of $A_{ij}$ when $\ket{b_i}\in {\cal H}_{1}$ and $\ket{b_j}\in {\cal H}_{2}$ is relaxed. In other words, one also allows $\bra{b_i}A\ket{b_j}\ne0$, and only one (trivial) zero eigenvalue remains, $\lambda_1=0$. However, it can be shown \cite{Newman2006-2} that the lowest nonzero eigenvalues of $L$ still encode the structure of the underlying graph. The latter stems from the fact that Eq.~\eqref{fiedler} can be viewed as a strategy for finding a minimal cut via connections/hops. This can be easily shown for a case with $\dim({\cal H}_{1})=\dim({\cal H}_{2})=\dim{\cal H}/2$. Using the Rayleigh variational principle, the smallest nonzero eigenvalue $\lambda_2$ of the Laplacian admits an upper bound obtained by evaluating the Rayleigh quotient, $\lambda_2 \le \,\mathbf{v}^T L \mathbf{v}/\mathbf{v}^T\mathbf{v}$. The test vector $\mathbf{v}$ should contain the same number of elements $v_i=1$ and $v_i=-1$ so that it is orthogonal to the eigenvector corresponding to $\lambda_1=0$. Using Eq.~\eqref{fiedler} one finds 
\begin{equation}
\lambda_2\leq {\cal R}=\frac{4}{\mathrm{dim}({\cal H}_1)+\mathrm{dim}({\cal H}_2)}\sum_{i\in {\cal H}_{1},j\in {\cal H}_{2}} A_{ij}\,,
\label{eq:l1}
\end{equation}
showing that the total weight of inter-subspace couplings controls the smallest nonzero eigenvalue. The latter reflects near fragmentation and can happen when ${\cal R}\ll1$, i.e., (i) when (initially) fragmented subspaces are joined only by a few connections, or (ii) when the inter-subspace connections are weak in comparison to intra-connections (i.e., in the case of the so-called {\it weighted graphs}). Below, we will discuss both of these cases. Adjacency matrices for the cases (i) and (ii) are shown in Fig.~\ref{fig1}(c) and Fig.~\ref{fig1}(d), respectively. Consequently, Eq.~\eqref{eq:l1} bounds the eigenvalue $\lambda_2$ calculated from the Laplacian matrix for a nearly fragmented system with ${\cal R} \ne 0$. In the fully fragmented case, one obtains $\lambda_2=0$. Consequently, ${\cal R}$ serves as a rough estimate for the changes in the smallest $\lambda_i$, which occur upon introducing a perturbation that breaks strict fragmentation.

\subsection{Modularity}
\label{sec:modularity}
It is important to note that, in contrast to fully fragmented systems -- where the number of zero eigenvalues of the Laplacian gives the exact number of fragmented subspaces -- in the nearly fragmented case, no such direct information is available. The eigenvalues of such $L$ are increasing, $\lambda_1<\lambda_2<\lambda_3<\dots$, and the bottom Laplacian eigenvalues often form a cluster rather than exhibiting a clear gap \cite{Davis1970,Luxburg2007,Luxburg2008,Yu2014}. Small perturbations (inter-subspace connections) can cause large rotations of eigenvectors, and there is no well-defined condition to stop the analysis. Consequently, the question of how many Fiedler vectors to investigate is inherently heuristic and model- and perturbation-dependent. As we will show below, another graph-theoretic measure, the so-called {\it modularity} \cite{Clauset2004,Newman2004,Newman2006-1}, is better suited to the analysis of nearly fragmented systems provided that the nearly decoupled subspaces are of comparable dimensions.

Instead of looking at the minimal cut through connections, one can focus on the comparison between the actual structure of the graph encoded in the adjacency matrix $A$ and the test graph with edges (connections) placed at random between vertices characterized by their degrees, $d_i=\sum_{j}A_{ij} $. The expected number of edges between two vertices of degree $d_i$ and $d_j$ is $d_i d_j/2m $, where $m=\sum_i d_i/2$. Then, one introduces a modularity matrix
\begin{equation}
Q_{ij}=A_{ij} - \frac{d_i d_j}{2m}\;,
\label{eq:mod}
\end{equation}
and the intuition here is as follows: since in a fragmented or nearly fragmented system, the number of edges within the subspace (between the subspaces) is larger (smaller) than average, one expects that cases with $Q_{ij}>0$ ($Q_{ij}<0$) represent $i\leftrightarrow j$ connection within the subspace (between the subspaces). Consequently, the strategy is to find a division of states (vertices) into two subspaces, ${\cal H}_1$ and ${\cal H}_2$, for which $\sum_{i,j \in {\cal H}_1 }Q_{ij}+\sum_{i,j \in {\cal H}_2}Q_{ij}$ is maximal. This optimization problem can also be reduced to the solution of the eigenproblem for the modularity matrix
\begin{equation}
\frac{\mathbf{v}^T Q \mathbf{v}}{\mathbf{v}^T \mathbf{v}} = \beta, 
\label{modularity}
\end{equation}
in a similar way as it is done for the Laplacian matrix, see Appendix~\ref{sec:ap1}. Note that there are important differences between the analysis of the Laplacian $L$ and modularity $Q$. In the former, we focus on the smallest eigenvalues; in the latter, we focus on the largest (i.e., on the best division of the graph). Consequently, in the following we will present the eigenvalues in the reversed order, $\widetilde{\beta}_i=\beta_{\mathrm{dim}({\cal H})}-\beta_{\mathrm{dim}({\cal H})+1-i}$. In such notation, the smallest  $\widetilde{\beta}_1<\widetilde{\beta}_2<\widetilde{\beta}_3<\dots$ correspond to the largest eigenvalues of $Q$. The other important difference concerns the trivial solution, $\mathbf{v}^T_1=(1,1,1,\dots)$, of the minimization problem from Eq.~\eqref{fiedler}. Such a vector does not occur if the partitioning is formulated in terms of modularity, since $\mathbf{v}^T_1 Q \mathbf{v}_1=0 < \beta_{\mathrm{dim}({\cal H})}$. Consequently, in the case of perfect fragmentation with $F$ subspaces, one needs to determine numerically either $F$ smallest eigenvalues and eigenvectors of the Laplacian or $F-1$ largest eigenvalues and eigenvectors of the modularity matrix \cite{Newman2006-1}. Note that also in the case of the Laplacian $L$, there are $F-1$ meaningful eigenvalues, since $\lambda_1$ corresponds to a trivial (constant) $\mathbf{v}_1$.

Finally, one can estimate a bound on the changes in the eigenvalues of $Q$ upon fragmentation-breaking perturbation, $\sum_{i\in {\cal H}_{1},j\in {\cal H}_{2}} A_{ij}$ [similar to Eq.~\eqref{eq:l1}]. In Appendix~\ref{sec:ap2}, we show that the latter yields $\beta - \beta' \le {\cal R}/2$, where $\beta$ and $\beta'$ denote the eigenvalues of the fragmented and perturbed systems, respectively. It is evident that for ${\cal R}\ll 1$, the system is again expected to be nearly fragmented.

\section{Nearly-fragmented $t$-$J$ model}
\label{sec:results}
We will now investigate the fragmentation of the Hilbert space using the Laplacian $L$ and the modularity $Q$. We will analyze two types of systems [see Fig.~\ref{fig1}(c,d)]. (i) In the first case, the subspaces are linked by only a few connections, ${\cal R}\ll1$. We achieve this by investigating the rung-depleted $t$-$J_{z}$ ladder: a two-leg ladder system in which selected rung bonds are removed \cite{Lisiecki2025}. In such a system, the holes hop without alternating the spin configuration in the region without rungs. On the other hand, in regions with rungs, hole hopping can alter the spin configuration. Such a mixture provides a controlled realization of a nearly fragmented system. (ii) The second system reflects the case with an overall large number of intra-subspace connections that are small, i.e., ${\cal R}\ll 1$. We achieve this by considering the full $t$-$J$ model with $t\gg J$. In such a case, the spin-flip term, $(S^{+}_{i}S^{-}_{i+1}+S^{-}_{i}S^{+}_{i+1})$, heavily mixes all spin configurations, but the weight of such connections is controlled by $J$, yielding full Hilbert space fragmentation of $t$-$J_{z}$ model in the $J=0$ limit. 

\begin{figure}[!t]
\includegraphics[width=1.0\columnwidth]{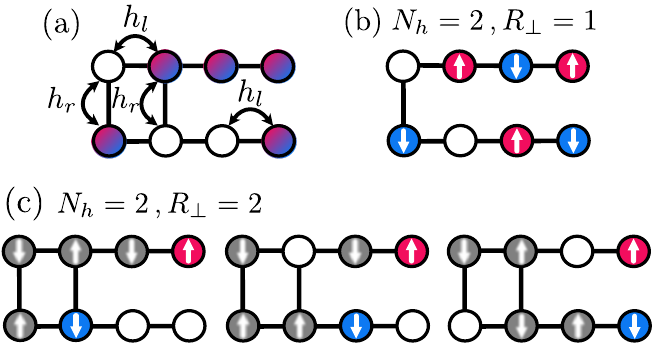}
\caption{Rung-depleted $t$-$J_{z}$ ladder. (a) Sketch of the system with $R_\perp=2$. (b) Exemplary spin configuration for $L=8$, $N_h=2$, and $R_\perp=1$. (c) Three exemplary configurations for $L=8$, $N_h=2$, and $R_\perp=2$. "Fixed" $q=2$ spins (which cannot be changed by the hole hop) are marked with red and blue colors.}
\label{sketch}
\end{figure}

\subsection{Rung-depleted $t$-$J_{z}$ ladder}
In this section, we discuss the graph theory analysis of the fragmentation within the rung-depleted $t$-$J_{z}$ ladder introduced in Ref.~\cite{Lisiecki2025}. The Hamiltonian of an $L$-site system is given by
\begin{equation}
H=\sum_{l=1}^{L/2-1} h_l(l) + \sum_{l=L/2+1}^{L-1} h_l(l) + \sum_{r=1}^{R_\perp} h_r(r)\,,\\
\label{eq:hamrdl}
\end{equation}
with
\begin{eqnarray}
h_l(l)&=&t \sum_{\sigma}\left(P_G\,c^\dagger_{l,\sigma}c^{\phantom{\dagger}}_{l+1,\sigma}P_G+\mathrm{H.c.}\right)\,,\\
h_r(r)&=&t \sum_{\sigma}\left(P_G\,c^\dagger_{L/2+1-r,\sigma}c^{\phantom{\dagger}}_{ {L/2+r,\sigma}}P_G\,+\mathrm{H.c.}\right)\,.
\end{eqnarray}
Here, the $h_l(l)$ represents a standard $t$-$J_{z}$ chain term, while the $h_r(r)$ denotes hopping on the rungs of the ladder structure [see sketch of the system presented in Fig.~\ref{sketch}(a)]. For $R_\perp=L/2$, the system realizes a full ladder geometry, while for $R_\perp=1$, it represents a \mbox{$t$-$J_{z}$} chain. 

It has been shown~\cite{Lisiecki2025} that depending on the number of holes, $N_h$, and rungs, $R_{\perp}$, the above system allows for sequential control over the number of fragmented subspaces. Consider the setup presented in Fig.~\ref{sketch}(b,c), i.e., $L=8$ and $N_h=2$. For $R_\perp=1$ [i.e., for a single chain shown in Fig. \ref{sketch}(b)], and $S^z_\mathrm{tot}=\sum_i S^z_i=0$ sector, the Hilbert space is fragmented into $20$ subspaces (reflecting configurations in which $L-N_h=6$ spins can be arranged in zero magnetization sector). Now, let's consider $R_\perp=2$ from Fig.~\ref{sketch}(c). In the region with rungs, the presence of holes can alter the overall spin configuration. However, the spins of the last two "outer" particles (far from the rung-rich region) cannot be changed. Consequently, the system will exhibit fragmentation into four subspaces. In the general $L\,,N_h\,,R_\perp$ case, the configuration of $q=L-2(N_h+R_\perp-1)$ spins will remain intact ("frozen"), resulting in 
\begin{equation}
F=\sum_{n=n_\mathrm{min}}^{n_\mathrm{max}} \binom{q}{n}\,,
\label{eq:numf}
\end{equation}
subspaces. Here \mbox{$n_\mathrm{min}=\mathrm{max}(0,q-N_{-\sigma})$}, \mbox{$n_\mathrm{max}=\mathrm{min}(q,N_{\sigma})$}, and $N_{-\sigma}\geq N_{\sigma}$ is the number of particles with given spin $\sigma$. Note that Eq.~\eqref{eq:numf} holds for $R_\perp>1$ and $q \ge 0$. For $R_\perp=1$, the total number of subspaces is given by the total number of spin configurations, $F=\binom{L-N_h}{N_{\sigma}}$. For $R_\perp=L/2$ (the full ladder geometry), one obtains $q<0$, so there is no fragmentation and $F=1$.

\subsubsection*{Eigenvalue analysis: number of subspaces}
\begin{figure}[!t]
\includegraphics[width=1.0\columnwidth]{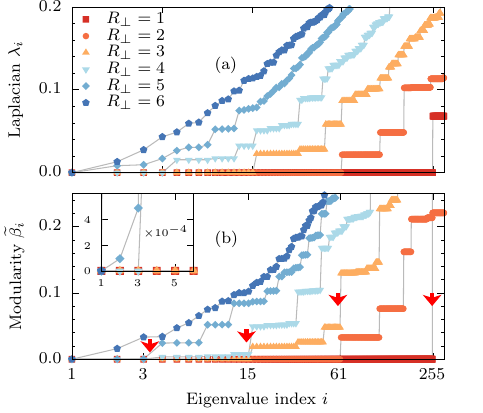}
\includegraphics[width=0.9\columnwidth]{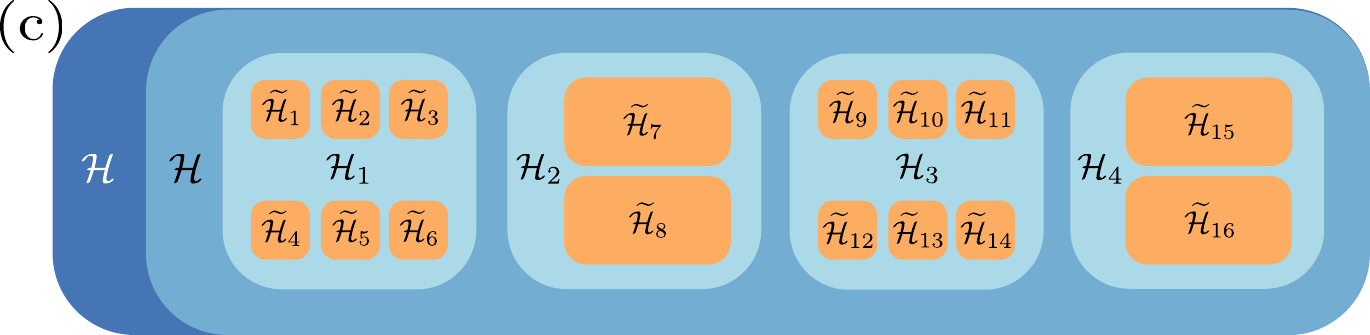}
\caption{Spectral graph analysis of rung-depleted $t$-$J_{z}$ ladder ($L=12$, $N_h=2$) for various numbers of rungs $R_\perp=1,\dots,6$. Panel (a) depicts the first $300$ eigenvalues $\lambda_i$ of the Laplacian $L$, while panel (b) depicts the last $300$ eigenvalues of modularity $Q$. In the latter, we present eigenvalues in the reversed order, $\widetilde{\beta}_i=\beta_{\mathrm{dim}(\mathcal{H})}-\beta_{\mathrm{dim}(\mathcal{H})+1-i}$. Note the log-scale on the $x$-axis. The inset in (b) shows a zoomed-in view of the first few modularity eigenvalues. (c) Sketch of the structure of the Hilbert space for $R_\perp=6,5,4,3$ ($F=1,1,4,16$, respectively).}
\label{fig3}
\end{figure}

In Fig.~\ref{fig3}, we present the spectral graph-theory analysis of the system defined above for $L=12$, $N_h=2$ (total Hilbert space dimension $\dim({\cal H})=16632$), and various $R_\perp=1,\dots,6$. The number of subspaces $F$ and "frozen" spins $q$ is given in Tab.~\ref{tab:blocks}. The analysis of eigenvalues of the Laplacian $\lambda$ and modularity $\beta$ of such a system is presented in Fig.~\ref{fig3}(a) and Fig.~\ref{fig3}(b), respectively. It is evident that both approaches properly describe the number of fragmented subspaces $F$.

\begin{table}[b!]
\centering
\begin{tabular*}{\columnwidth}{@{\extracolsep{\fill}}|c|c|c|c|c|c|c|@{}}
\hline
$R_\perp$ & 1   & 2  & 3  & 4 & 5 & 6 \\ \hline
$q$       & 10  & 6  & 4  & 2 & 0 & 0 \\ \hline
$F$       & 256 & 62 & 16 & 4 & 1 & 1 \\ \hline
\end{tabular*}
\caption{The number of subspaces $F$ and "frozen" spins $q$ versus number of rungs $R_\perp$. Here we consider $L=12$ and $N_h=2$.}
\label{tab:blocks}
\end{table}

Interestingly, although for all consider cases \mbox{$F=\dim(\lambda_i=0)$} or \mbox{$F=\dim(\widetilde{\beta}_i=0)+1$}, for the first lattice configuration for which fragmentation is absent, i.e., for $R_\perp=5$ with $F=1$, the largest eigenvalues of the modularity $Q$ are close to being degenerated [see inset in Fig.~\ref{fig3}(b)]. In fact, there is the whole cascade of "jumps"/gaps [marked with red arrows in Fig.~\ref{fig3}(b)] in eigenvalues of modularity, indicating a clustering of $\beta_i$ around some values. The latter arises from the fact that the system with a smaller number of rungs, i.e, the $R_\perp-1$ system, exhibits fragmentation. Adding an extra rung to the $R_\perp-1$ system introduces rare connections between subspaces so that the $R_\perp$ system no longer exhibits strict fragmentation. However, as we show below, the clustering of $\beta_i$  in the $R_\perp$ system is a signature of a nearly-fragmented Hamiltonian with ${\cal R}\ll1$ in Eq.~\eqref{eq:l1} and of the presence of $nF$ weakly coupled communities. Throughout this work, we denote by $F$ the number of strictly decoupled subspaces, and by $nF$ the number of weakly coupled communities. The spectrum of modularity contrasts with the behavior of the Laplacian eigenvalues, which approach zero continuously in cases with $F=1$ (i.e., for $R_\perp=5$ and $R_\perp=6$), so they do not reveal any clear signature of a nearly fragmented system.

\begin{figure}[!t]
\includegraphics[width=1.0\columnwidth]{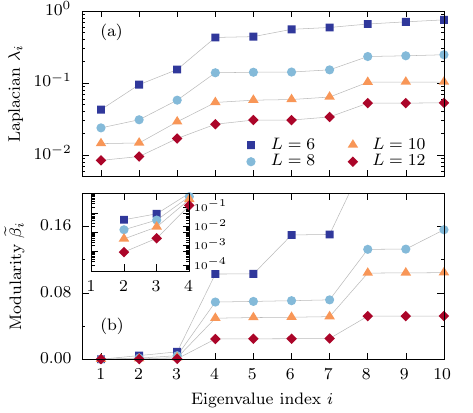}
\caption{Lowest eigenvalues (a) $\lambda_i$ of the Laplacian $L$ and (b) $\widetilde{\beta}_i=\beta_{\mathrm{dim}(\mathcal{H})}-\beta_{\mathrm{dim}(\mathcal{H})+1-i}$ of the modularity $Q$, for various system sizes $L=6,8,10,12$ and $N_h=2$ [with $\dim({\cal H})=90,560,3150,16632$, respectively] and $R_\perp=L/2-1$. Note that all considered systems have $F=1$. The inset in (b) shows a zoomed-in view of the first few modularity eigenvalues.}
\label{fig4}
\end{figure}

\begin{figure*}[!t]
\includegraphics[width=1.0\textwidth]{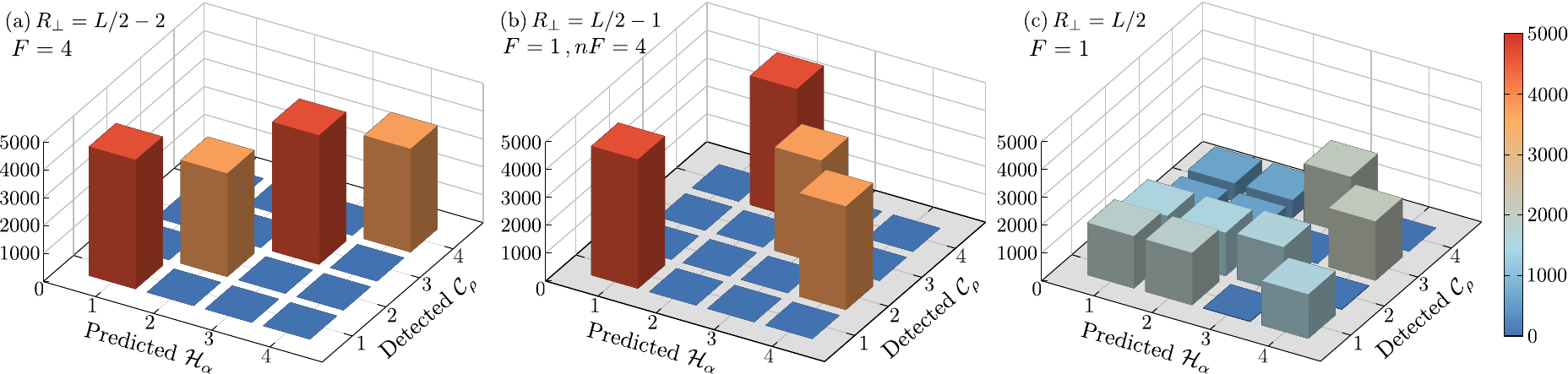}
\caption{Subspace detection based on $k$-means algorithm (see text for details). Panel (a) illustrates the starting point, i.e., the strictly fragmented system ($F=4$\,, $R_\perp=L/2-2=4$)  for that we put ${\cal H}_{\alpha}={\cal C}_{\alpha}$, (b) the nearly fragmented system ($F=1\,, nF=4\,,R_\perp=L/2-1=5$), and (c) a system for which we don’t see any structure in the spectral analysis of the corresponding graph. In all panels, $L=12$ and $N_h=2$. See also Fig.~\ref{fig3} for eigenvalues of $L$ and $Q$. The plots depict two-dimensional histograms of $(\alpha,\rho)$ pairs, where $\alpha$ index corresponds to the theoretically predicted ${\cal H}_{\alpha}$ subspace, while $\rho$ corresponds to the detected subspace by the $k$-means algorithm based on eigenvectors of modularity $Q$. See text for details.}
\label{fig5}
\end{figure*}

Consider the setups in which fragmentation is absent ($F=1$) but the number of additional rungs, $R_\perp$, is chosen so that the $R_\perp-1$ system is fragmented. Note that varying $R_\perp$ in~\eqref{eq:hamrdl} results in sequentially joining the subspaces, i.e., ${\cal H}_\beta(R_\perp)=\bigoplus_{\alpha} {\cal H}_{\alpha}(R_\perp-1)$ [see also sketch presented in Fig.~\ref{fig3}(c)] . Here, the direct sum goes only over subspaces that have the same frozen spins at the edges of the system, far from the rung-rich region (the number of such spins depends on the specific value of $R_\perp$). Below, we analyze the case with $N_h=2$ holes, which, for $R_\perp=L/2-1$, yields $F=1$, while for $R_\perp=L/2-2$ gives $F=4$ (for all system sizes). We aim to detect the existence and, potentially, reveal the structure of the underlying subspaces of ${\cal H}$, even though there is no strict fragmentation, i.e., with $F=1$. As mentioned in Sec.~\ref{sec:gt}, the main issue in finding the underlying community (subspace) structure is determining the appropriate number of eigenvectors of $L$ or $Q$ to analyze. In Fig.~\ref{fig4}, we again turn to the eigenvalues of Laplacian $L$ and modularity $Q$ of systems of various sizes $L=6,8,10,12$, $N_h=2$, and $R_\perp=L/2-1$, all with $F=1$. As mentioned previously, the Laplacian eigenvalues $\lambda_i$ don't reveal any structure, i.e., don't have any well-developed gaps $\lambda_{i+1}-\lambda_{i}$ [see Fig.~\ref{fig3}(a) for $R_\perp=1,2$ and Fig.~\ref{fig4}(a) for all presented $R_\perp$]. However, the behavior of the modularity $Q$ is different: for all considered systems, the gaps in the latter, $\widetilde{\beta}_{i+1}-\widetilde{\beta}_{i}$, can be clearly seen in Fig.~\ref{fig4}(b). More importantly, all cases have $\mathrm{dim}(\widetilde{\beta}_i<\eta)=3$, with $\eta\to0$ with increasing system size or, more specifically, with increasing $\mathrm{dim}({\cal H})$ [see inset of Fig.~\ref{fig4}(b)]. We argue that such behavior is a manifestation of nearly fragmented systems with
\begin{equation}
nF=\mathrm{dim}(\widetilde{\beta}_i<\eta)+1
\label{eq:nf}
\end{equation}
communities that are weakly connected [in the sense that ${\cal R}\ll 1$]. Such a scenario is depicted also in Fig.~\ref{fig1}(c) for $L=6\,,N_h=2$ and $R_\perp=L/2-1=2$.

\subsubsection*{Eigenvector analysis: community detection}
The multiplicity of the zeros of the $\lambda_i$ or $\widetilde{\beta}_i$ eigenvalues (or their proximity to zero) provides information about the number of communities ($F$ for strict fragmentation or $nF$ for nearly fragmented system). On the other hand, to find the community itself within the Hilbert space ${\cal H}$ (i.e., to assign the basis states $\ket{b_i}$ to subspaces ${\cal H}_{\alpha}$), we have to analyze the eigenvectors of $L$ or $Q$. The numerical solution of the eigenproblem in Eq.~\eqref{fiedler} or Eq.~\eqref{modularity} gives arbitrarily rotated and normalized eigenvectors $\mathbf{u}^{\alpha}$, where $(\mathbf{u}^{\alpha})^T=(u^{\alpha}_1,u^{\alpha}_2,u^{\alpha}_3,\dots,u^{\alpha}_{\mathrm{dim}({\cal H})})$. However, the latter rotation preserves an important property for fully fragmented systems ($F>1$): for every pair of basis states $\ket{b_i}$ and $\ket{b_j}$ that belong to the same ${\cal H}_{\alpha}$ all components are equal, i.e., $u^{\alpha}_i=u^{\alpha}_j$ for every $\alpha=1,\dots,F$. Using the numerical eigenvectors, $\mathbf{u}^{\alpha}$, we introduce for each state $\ket{b_i}$ an $F$-dimensional vector $\mathbf{R}_i=(u^{1}_i,u^{2}_i,u^{3}_i,\dots,u^F_i)$. Then, states with the same $\mathbf{R}_i$ vectors belong to the same subspace ${\cal H}_{\alpha}$. 

For the nearly-fragmented systems ($F=1$ and $nF>1$), the states $\ket{b_i} $ can be grouped into $nF$ distinct communities ${\cal C}_{\alpha}$ according to the distances $|\mathbf{R}_i-\mathbf{R}_j|$, for which we apply the $k$-means algorithm. First we find two states $\ket{1}$ and $\ket{2}$ for which $|\mathbf{R}_1-\mathbf{R}_2|$ is maximal and we assign them to ${\cal C}_{1}$ and ${\cal C}_{2}$, respectively. Then we find a third state $\ket{3} \in {\cal C}_{3}$ that maximizes $|\mathbf{R}_1-\mathbf{R}_3|^2+|\mathbf{R}_2-\mathbf{R}_3|^2$ and continue until $nF$ distant states, $\ket{\alpha}$, $\alpha=1,\dots,nF$ are assigned to $nF$ communities ${\cal C}_{\alpha}$. The remaining $\mathrm{dim}({\cal H})-nF$ states are assigned according to the shortest distance from the current centroid of the given ${\cal C}_\alpha$. Next, new centroids are calculated, and the assignments are then iteratively updated till convergence. It is important to note that the number $C$ of desired communities ${\cal C}_\alpha$ is a free parameter of the $k$-means algorithm. Consequently, recognizing that $C=nF$, as obtained through eigenvalue analysis, is of great importance.

Let's consider the configuration presented in Fig.~\ref{fig3}, i.e., $L=12$, $N_h=2$, and cases $R_\perp=4,5,6$ representing the fragmented system with $F=4$, the nearly fragmented system with $F=1\,, nF=\mathrm{dim}(\widetilde{\beta}_{i}<\eta=0.01)+1=4$, and a system for which we don't see any structure in eigenvalues of $L$ or $Q$, respectively. To show that the above algorithm correctly identifies the communities in the nearly fragmented system, we first assign each basis state $\ket{b_i}$ (vertex) to the theoretically predicted subspaces ${\cal H}_{\alpha}$. This can be achieved, e.g., by grouping the $\mathbf{R}_i$ vectors for the $R_\perp=L/2-2=4$ case, $F=4$, as described above. Here we build $\mathbf{R}_i$ based on eigenvectors corresponding to $\mathrm{dim}(\widetilde{\beta}_{j}<\eta=0.01)+1=4$.  Next, we apply the $k$-means algorithm for a larger $R_\perp$, which assigns the communities ${\cal C}_\rho$ in a system that does not show strict fragmentation. Consequently, each $\ket{b_i}$ is labeled by a pair of indices $(\alpha,\rho)$, where the first one represents the subspace in a strictly fragmented system and the second one is the community identified by $k$-means in a nearly fragmented case. If all basis states are properly assigned in a nearly fragmented system ($F=1\,, nF>1$), then the two-dimensional histogram of pairs $(\alpha,\rho)$ is a monomial matrix (generalized permutation matrix with one nonzero element in each row and column). The latter stems from the fact that all basis states from the ${\cal H}_{\alpha}$ subspace are assigned to the same ${\cal C}_\rho$ community. Such case is presented in Fig.~\ref{fig5}(b) for $F=1\,,nF=4$ ($R_\perp=5$). On the other hand, trying to find $4$ communities for the $R=L/2=6$ case (the setup without any structure in $\lambda$ or $\beta$) fails and histogram of $(\alpha,\rho)$ has many nonzero elements in each raw/column [see Fig.~\ref{fig5}(c)], indicating that the states from some (fixed) ${\cal H}_{\alpha}$ are assigned to various ${\cal C}_\rho$.

\subsubsection*{Simplied approach to clustering}

Finally, we want to point out that in some cases, knowledge of $F$ or $nF$ and of the eigenvectors of $Q$ is sufficient to identify the subspaces, without involving the $k$-means algorithm. Consider again the systems presented in Fig.~\ref{fig4} for which modularity eigenvalues yield $nF=\mathrm{dim}(\widetilde{\beta}_{j}<\eta=0.01)+1=4$ and its two nontrivial numerically obtained eigenvectors $Q\mathbf{u}_{j}=\beta_{j}\mathbf{u}_{j}$, with $\mathbf{u}_{\mathrm{dim}({\cal H})}=\sum_i \Lambda_i\ket{b_i}$ and $\mathbf{u}_{\mathrm{dim}({\cal H})-1}=\sum_i \Gamma_i\ket{b_i}$. Our results indicate that a simple analysis of $\Lambda_i$ and $\Gamma_i$ of such eigenvectors is sufficient to detect the subspaces. We find that the following clustering
\begin{eqnarray}
{\cal C}_1 &=& \mathrm{span}
\big\{\ket{b_i} \;:\; \Lambda_i>\zeta\;
,\; |\Gamma_i|\leq\zeta \big\}\,,\nonumber\\
{\cal C}_2 &=& \mathrm{span}\big\{\ket{b_i} \;:\; \Lambda_i<\zeta\;
,\; |\Gamma_i|\leq\zeta \big\}\,,\nonumber\\
{\cal C}_3 &=& \mathrm{span}\big\{\ket{b_i} \;:\; \Gamma_i>\zeta\;
,\; |\Lambda_i|\leq\zeta \big\}\,,\nonumber\\
{\cal C}_4 &=& \mathrm{span}\big\{\ket{b_i} \;:\; \Gamma_i<\zeta\;
,\; |\Lambda_i|\leq\zeta\big\}\,,
\label{eq:clust}
\end{eqnarray}
with $\zeta=1/\left(L\sqrt{\mathrm{dim}(\cal H)}\right)$ correctly identifies all ${\cal C}_{\alpha=1,2,3,4}$. Note that the above construction is a consequence of ideal (not rotated) eigenvectors of $L$ discussed in Sec.~\ref{sec:gt}. 

\begin{figure*}[!t]
\includegraphics[width=1.0\textwidth]{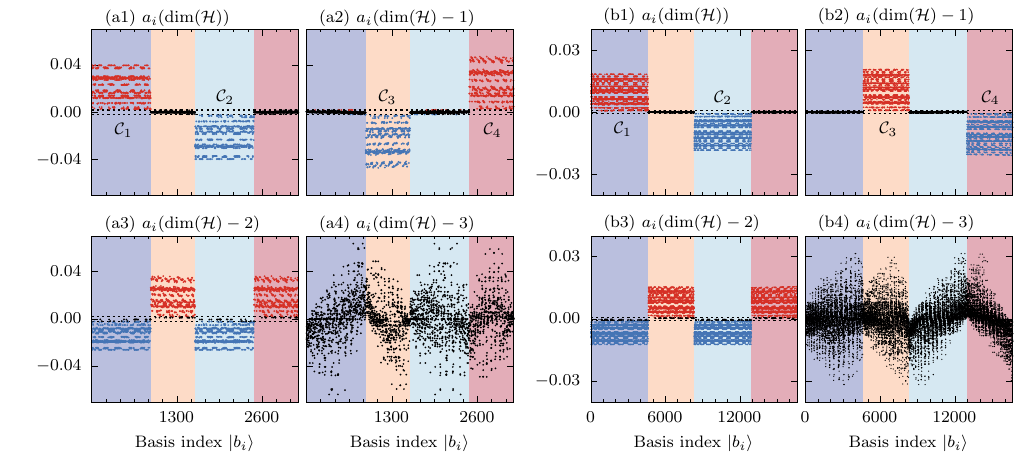}
\caption{Community (subspace) detection based on the eigenvectors $\mathbf{u}$ of the modularity $Q$. The basis states $\ket{b_i}$ (the $x$-axis) are sorted in such a way that they belong to subsequent subspaces ${\cal H}_i$ (marked with shaded regions in all panels). See the text for the details. Left (a) and right (b) panels represent results for $L=10$ and $L=12$, respectively, both with $N_h=2$ and $R_\perp=L/2-1$. In both panels we present the coefficients $a_i(j)$ introduced in Eq.~\eqref{coeff} of the eigenvectors $\mathbf{u}_j$ corresponding to the largest eigenvalues $j=\mathrm{dim}({\cal H}),\mathrm{dim}({\cal H})-1,\mathrm{dim}({\cal H})-2,\mathrm{dim}({\cal H})-3$.}
\label{fig6}
\end{figure*}

In Fig.~\ref{fig6} we demonstrate the above clustering, Eq.~\eqref{eq:clust}, for $N_h=2$ and $L=10,12$ sites. Firstly, we establish the exact subspaces ${\cal H}_{\alpha=1,2,3,4}$ by again solving $R_\perp=L/2-2$ (yielding $F=4$, see Fig.~\ref{fig3}). This allows us to sort the basis states $\ket{b_i}$ (the $x$-axis in Fig.~\ref{fig6}) in such a way that they belong to subsequent subspaces ${\cal H}_{\alpha}$, e.g.,
\begin{eqnarray}
\mathbf{u}_{j}&=&
\sum_{i=1}^{Z_1} a_i(j)\ket{b_i}+
\sum_{i=Z_1+1}^{Z_1+Z_2}a_i(j)\ket{b_i}\nonumber\\
&+&
\sum_{i=Z_1+Z_2+1}^{Z_1+Z_2+Z_3}a_i(j)\ket{b_i}+
\sum_{i=Z_1+Z_2+Z_3+1}^{Z_1+Z_2+Z_3+Z_4}a_i(j) \ket{b_i}\,, \nonumber \\
\label{coeff}
\end{eqnarray}
where $Z_{\alpha}=\mathrm{dim}({\cal H}_{\alpha})$. Next, we present coefficients of the last four eigenvectors $\mathbf{u}_{j}$ of $Q$ obtained for a nearly fragmented system with  $R_\perp=L/2-1$, $F=1$, and $nF=4$. It is evident that the found subspaces ${\cal C}_\alpha$ have perfect overlap with the predicted ${\cal H}_\alpha$. Furthermore, in the same figure we present the coefficients of the first eigenvector corresponding to eigenvalue that does not fulfill the $\widetilde{\beta}_{j}<\eta$ condition, i.e., $\mathbf{u}_{\mathrm{dim}({\cal H})-3}$ [see Figs.~\ref{fig6}(a4) and \ref{fig6}(b4)]. As expected, in the latter case, we don't observe any fragmentation-related structure.

It should be pointed out that $\eta$ and $\zeta$ thresholds in Eq.~\eqref{eq:nf} and Eq.~\eqref{eq:clust}, respectively, are, in principle, free parameters in our considerations. However, our results indicate that both of them decrease with increasing $\mathrm{dim}({\cal H})=\sum_\alpha \mathrm{dim}({\cal H}_\alpha)$. Furthermore, the $\eta$ value can be easily determined from the eigenvalues $\beta_j$, and its exact magnitude is not crucial (see Fig.~\ref{fig3} and Fig.~\ref{fig4}). Finally, the usage of the unknown $\zeta$ thresholds can be omitted by using the $k$-means clustering. The latter method is more general and can be applied to more complex cases that the simplified method discussed in this subsection does not properly capture.

\subsection{Full $t$-$J$ model}
\label{sec:gtj}
In the previous section, we have investigated the system in which one has direct control over the number of subspaces $F=F(R_\perp)$. As a consequence, the nearly fragmented system appears "at the end" of the sequence of fragmented ladders (for several rungs close to the maximal possible $R_\perp=L/2$) and is "built" out of a small number of communities ($nF=4$). In this section, we will consider the much harder problem of the full $t$-$J$ chain~\eqref{eq:tj} in which $nF$ can be exponentially large.

In the limit of $H_J=0$, the $t$-$J$ model exhibits full Hilbert space fragmentation into $F=\binom{L-N_h}{(L-N_h)/2}$ subspaces, where the only possible dynamics is induced by hopping $t$ within subspaces (in the following we call it {\it fast dynamics}). Introducing any finite $J\ne0$ connects all of the latter, leading immediately to $F=1$, in which the dynamics is also possible by the spin flips $J$ (we call it {\it slow dynamics}). Note that for the $t$-$J$ model with a small number of holes considered here, the cardinality (number of nonzero elements) of the $H_J$ term in~\eqref{eq:tj} is typically larger than that of $H_t$ [see Fig.~\ref{fig1}(d) for a sketch of the connectivity]. However, ${\cal R}\propto \sum_{i\in {\cal H}_{i}\ne j\in {\cal H}_{j}} A_{ij}$ in Eq.~\eqref{eq:l1} is controlled by the connections between subspaces given by $J$. Consequently, for $J\ll t$ one can expect that ${\cal R}\ll 1$. Such a limit is directly relevant for the $t$-$J$ model, since the latter is usually derived from the Hubbard model in the limit of large interactions $U$, i.e., $t\ll U\propto 1/J$. In the following, we demonstrate that, for sufficiently small $J$, the system indeed becomes nearly fragmented into $nF=\binom{L-N_h}{(L-N_h)/2}$ subspaces inherited from the $J=0$ limit. 

It is worth noting that there is a conceptual difference between the $t$-$J_{z}$ and the $t$-$J$ models in the construction of the Laplacian $L$ and modularity $Q$. As already mentioned, the $L$ and $Q$ matrices are built based on the kinetic energy of the Hamiltonian. Consequently, for the $t$-$J_{z}$ model, the adjacency matrix $A$ elements take only one out of two values: $A_{ij}=0$ or $A_{ij}=1$ (e.g., by choosing $t=1$), and thus the matrix has the form of an {\it unweighted graph}. On the other hand, in the $t$-$J$ model, there are two kinetic terms: the electron hopping $t$ (which we again set to unity, $t=1$) and the spin-flip term $S^+_iS^-_j$ given by $J\ne t$. The graph corresponding to such an adjacency matrix is called a {\it weighted graph}. 

\begin{figure}[!t]
\includegraphics[width=0.95\columnwidth]{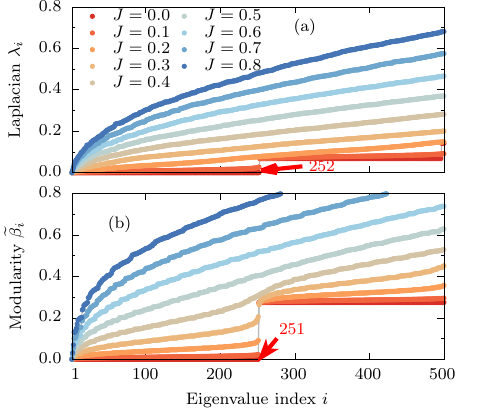}
\caption{Lowest eigenvalues of (a) Laplacian $L$ and (b) modularity $Q$ for the full $t$-$J$ model for various strengths of spin exchange $J=0.0\,,0.1\,,\dots\,,0.8$. Calculated for $L=12$ sites and $N_h=2$ holes, yielding $F=252$ in the fully fragmented limit.}
\label{fig7}
\end{figure}

\begin{figure}[!t]
\includegraphics[width=0.95\columnwidth]{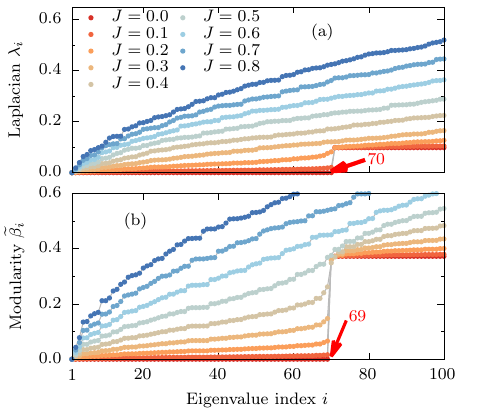}
\includegraphics[width=0.80\columnwidth]{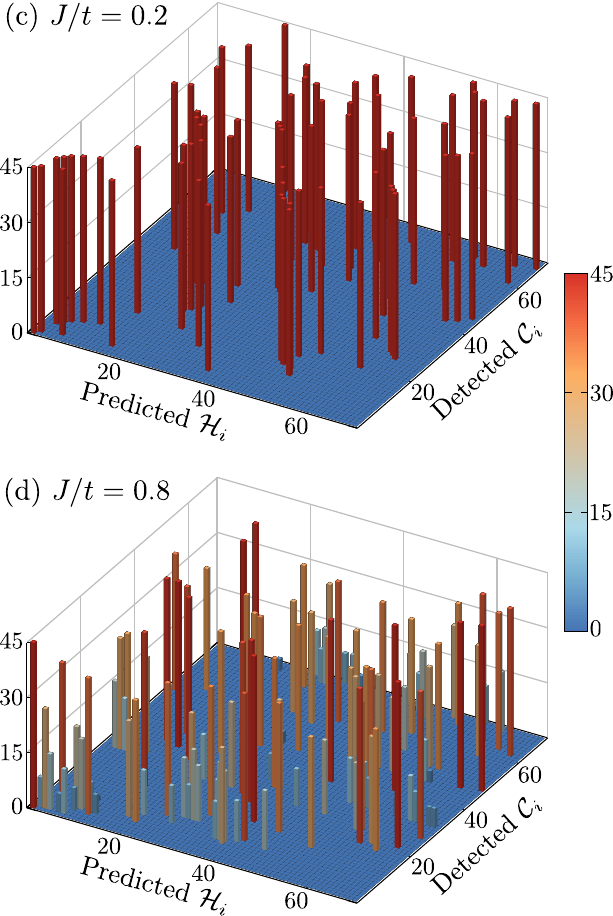}
\caption{(a,b) Similar analysis as in Fig.~\ref{fig7} but $L=10$, $N_h=2$, and $nF=70$. (c,d) Similar analysis as in Fig.~\ref{fig5} but for the full t-J model ($L=10\,,N_h=2$). The subspace detection is based on the eigenvectors $\mathbf{u}$ of the modularity $Q$  with (c) $J=0.2$ and (d) $J=0.8$. The plots depict two-dimensional histograms of $(\alpha,\rho)$ pairs, where $\alpha$ index corresponds to the theoretically predicted ${\cal H}_\alpha$ subspace of strictly fragmented model ($J=0$), while $\rho$ corresponds to the detected communities ${\cal C}_\alpha$ by the $k$-means algorithm based on eigenvectors of modularity $Q$.}
\label{fig8}
\end{figure}

In Fig.~\ref{fig7}, we examine the spectral properties of the $t$-$J$ graph for $L=12$, $N_h=2$, and several values of the coupling $J$. As in the rung-depleted ladder discussed in the previous section, the $\lambda$ and $\beta$ spectra of $L$ and $Q$, respectively, correctly recover the expected number of subspaces ${\cal H}_\alpha$ in the fully fragmented limit $J=0$ ($F=252$ for the parameters considered here). Our results further show that the $t$-$J$ model preserves its community structure up to $J/t\simeq {\cal O}(0.5)$, with $nF=F$. This is clearly visible in the modularity eigenvalues $\widetilde{\beta}_i$ shown in Fig.~\ref{fig7}(b), whereas the Laplacian eigenvalues $\lambda_i$ [Fig.~\ref{fig7}(a)] are more sensitive to the perturbation induced by finite $J$. Figure~\ref{fig8} presents the corresponding results for the $L=10$, $N_h=2$ system ($F=70$ in the $J=0$ limit), together with the reconstructed communities $C_i$ compared against the fully fragmented subspaces ${\cal H}_i$.

\begin{figure*}[!t]
\includegraphics[width=1.0\textwidth]{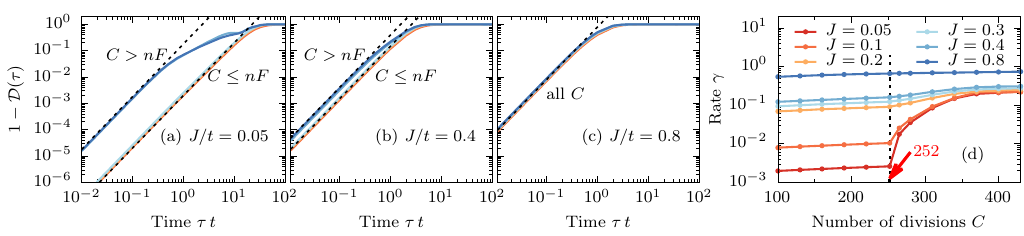}
\caption{Time evolution of the subspace projected Loschmidt echo, Eq.~\eqref{do2}, calculated for the full $t$-$J$ model with various value of spin exchange $J/t=0.05\,,0.4\,,0.8$ and various divisions into communities $C=130\,,190\,,252=F_0\,,340\,,400$. Note the log-log scale and black dashed lines indicating $\propto \tau^2$. For considered system, $L=12$ and $N_h=2$, the $U\to \infty$ yields $F_0=252$. See text for details. Panel (a,b) depicts results for nearly fragmented systems $J/t=0.05,0.4$ with ($nF=252$), while (c) shows results for a trivial ($nF=1$) system with $J/t=0.8$. (d) Dependence of $\gamma$ rate, as obtained from ${\cal D}(\tau)=1-\gamma\,\tau^2$ fit in the $\tau\leq 0.1/t$ interval, on the number of divisions $C$ within $k$-means algorithm. Calculated for $L=12$, $N_h=2$  and spin exchange $J=0.05\,,0.1\,,0.2\,,0.3\,,0.4\,,0.8$ (with $C=F_0=252$ in the $J=0$ limit).}
\label{fig9}
\end{figure*}

\section{Quantum dynamics vs. graph partitioning}
\label{sec:dynamics}
In this section, we will demonstrate that the subspaces ${\cal C}_{\alpha}\simeq{\cal H}_{\alpha}$ of nearly fragmented systems constructed via the graph decomposition are not mere mathematical artifacts but instead reflect the actual dynamics of the quantum system. To this end we consider time-evolution of the state $\ket{b_j} \in {\cal C}_{\alpha}$ and project it on other basis states $\ket{a_j}$ of the same subspace (or community) 
\begin{eqnarray}
{\cal D}_{\ket{b_j}}(\tau)& = &\sum_{\ket{a_j}\in {\cal C}_{\alpha}}
|\bra{a_j} e^{-iH\tau}\ket{b_j}|^2\,, \label{do1} \\
{\cal D}(\tau)&=&\frac{1}{Z} \sum_{\alpha} \sum_{\ket{b_j}\in {\cal C}_{\alpha}} {\cal D}_{\ket{b_j}}(\tau) \,, \label{do2} 
\end{eqnarray}
where $Z=\sum_\alpha\mathrm{dim}(C_\alpha)$ is the dimension of the Hilbert space. Note that within $k$-means algorithm, all $\ket{b_i}$ states are assigned to some ${\cal C}_{\alpha}$. The above equation represents a subspace projected Loschmidt echo. Consider a fully fragmented system: the dynamics remain completely confined within subspaces, and one will obtain ${\cal D}_{\ket{b_j}}(\tau)=\mathrm{const}=1$ if all states are correctly assigned to the proper communities, ${\cal C}_{\alpha}={\cal H}_{\alpha}$. On the other hand, if the states are wrongly assigned or the subspace structure given by ${\cal C}_{\alpha}$ does not reflect any physical processes in the dynamics, the ${\cal D}(\tau)$ rapidly approaches $0$. Consequently, the ${\cal D}(\tau)$ can capture not only the "bottlenecks" of the dynamics in nearly fragmented systems, but also can test whether the graph decomposition (discussed in the previous section) provides an adequate description of the actual quantum dynamics. It should be noted that for nearly fragmented systems with $F=1$, the long-time value of the subspace-projected Loschmidt echo vanishes, ${\cal D}(\tau\to\infty)\to0$, since such systems are ergodic.

In Fig.~\ref{fig9} we show the time-dependece, $1-{\cal D}(\tau)$, obtained for the full \mbox{$t$-$J$} model with $L=12$ sites and $N_h=2$ holes ($S^z_\mathrm{tot}=0$ magnetization sector). Fully fragmented system with $J=0$ has $F_0=252$ disconnected subspaces, ${\cal H}_{\alpha}$. For $J>0$ (with $F=1$), as a first step of the analysis, we detect $nF$ and assign to ${\cal C}_\alpha$ all of the basis states as described in Sec.~\ref{sec:gtj}. In what follows, we will use the eigenvalues and eigenvectors of the modularity matrix $Q$. We will also present results for $C\ne nF$ communities by requiring the $k$-means algorithm to partition into an arbitrary number $C$ of communities.

For all considered cases, the short-time dynamics is given by ${\cal D}(\tau\ll1)\simeq1-\gamma\,\tau^2$ and our results indicate that the rate $\gamma$ strongly depends on the considered setup. Let us first focus on $J/t=0.05$, i.e., the case for which spectral graph analysis yields $nF=F_0=70$ and ${\cal C}_{\alpha}={\cal H}_{\alpha}$. Our analysis indicates that [see Fig.~\ref{fig9}(a)] there is a few orders of magnitude difference between the values of $\gamma$ evaluated for $C\leq nF$ and $C>nF$. These results show how crucial is the knowledge of the $nF$ and $C_\alpha$ structures. In Fig.~\ref{fig9}(b), we also present a similar case for $J/t=0.4$. On the other hand, for $J/t=0.8$ [see Fig.~\ref{fig9}(c)], our results indicate that there is only one timescale, regardless of the Hilbert space division $C$. This is consistent with $\widetilde{\beta}_i$ eigenvalues analysis of such a system, which gives $F=nF=1$, i.e., the absence of subspaces (compare with Fig.~\ref{fig7} and Fig.~\ref{fig8})

The initial slow dynamics of ${\cal D}(\tau)$ for $C\leq nF$ can be interpreted as follows: starting from some state $\ket{b}\in C_\alpha$, the system first "explores" its community/subspace by {\it fast processes} governed by $t$ (i.e., the rearrangement of the holes) yielding ${\cal D}_{\ket{b_j}}(\tau)\simeq 1$. Next, the {\it slow dynamics} between communities takes place, controlled by the inter-subspace process $J$. Simple short time expansion of ${\cal D}(\tau)$ (see Appendix~\ref{sec:ap3})
\begin{eqnarray}
{\cal D}(\tau)&=& \frac{1}{Z} \sum_{\alpha} \Tr\left[\Pi_\alpha - \tau^2\,\Pi_\alpha H(\mathds{1}-\Pi_\alpha)H \right] \nonumber \\
&=&1-\frac{\tau^2}{Z}\sum_{\alpha \ne \rho} \sum_{i \in C_{\alpha},j \in C_{\rho}} |\bra{b_i}H \ket{b_j}|^2\;,\ \label{dtmain}
\end{eqnarray}
where $\Pi_\alpha = \sum_{\ket{b_i}\in {\cal C}_{\alpha}}\ket{b_i}\bra{b_i}$ is a projector into ${\cal C}_\alpha$ (with $\sum_\alpha \Pi_\alpha=\mathds{1}$), confirms this scenario. It is evident that to make any dynamics possible $(\mathds{1}-\Pi_\alpha)H\ket{b}\ne0$, which is possible only if the $J$ processes are present. Most importantly, Eq.~\eqref{dtmain} demonstrates that defining the adjacency matrix in terms of the squared Hamiltonian matrix, $A_{ij}=|\bra{b_i}H \ket{b_j}|^2$, establishes a direct connection between the graph weights and the quantum dynamics. The latter is evident since
\begin{equation}
    \frac{1}{Z}\sum_{\alpha \ne \rho} \sum_{i \in C_{\alpha},j \in C_{\rho}} |\bra{b_i}H \ket{b_j}|^2 = \frac{{\cal R}}{4}\,,
    \label{dtsecond}
\end{equation}
for nearly fragmented systems, yielding \mbox{${\cal D}(\tau)=1-\tau^2 \,{\cal R}/4$} for short times.

\begin{figure}[!t]
\includegraphics[width=1.0\columnwidth]{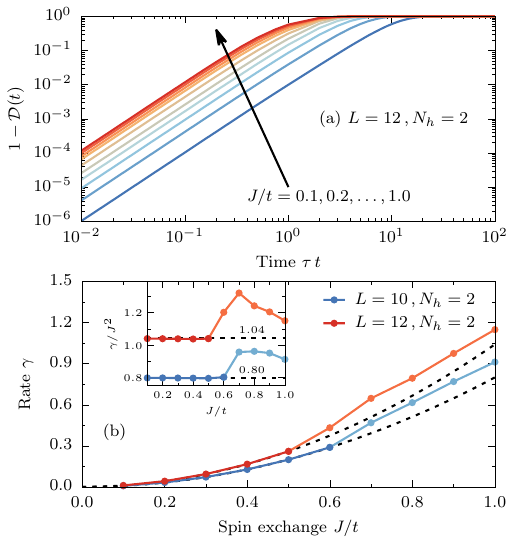}
\caption{(a) Time evolution of the subspace projected Loschmidt echo, Eq.~\eqref{do2}, calculated for the full $t$-$J$ model with various value of spin exchange $J/t=0.1\,,0.2\,,\dots\,,1.0$ and $L=12$, $N_h=2$ with $C=F_0=252$ communities. Note the log-log scale. (b) Dependence of $\gamma$ rate, as obtained from ${\cal D}(\tau)=1-\gamma\,\tau^2$ fit in the $\tau\leq 0.1/t$ interval, on the spin exchange $J$. Calculated for $L=10\,,N_h=2$ with $C=F_0=70$ and $L=12\,,N_h=2$ with $C=F_0=252$. The black dashed line represents the $J^2$ dependence, while the inset depicts $\gamma/J^2$ scaling.}
\label{fig10}
\end{figure}

In Fig.~\ref{fig10}(a) we show $1-{\cal D}(\tau)$ for various strength of spin exchange $J$ and $C=F_0$. Note that in this setup, we fix the number of divisions $C$ to the value obtained in the $J=0$ limit, while the communities ${\cal C}_\alpha$ are obtained from the $k$-means algorithm for a given $J\ne0$. Our results again indicate a strong dependence of ${\cal D}(\tau)$ on the parameters considered. In Fig.~\ref{fig10}(b) we present $\gamma$ rate dependence on $J$ as obtained from ${\cal D}(\tau)=1-\gamma\,\tau^2$ fit in the $\tau\leq 0.1/t$ interval. We find that for $J/t\lesssim0.5$ the rate $\gamma\propto J^2$, while for $J/t\gtrsim 0.5$ we observe clear deviations from this dependence. Note that this change in the behavior of $\gamma$ upon varying $J$ coincides with the changes observed in the spectra presented in Fig.~\ref{fig7}(b) and Fig.~\ref{fig8}(b). We again argue that the quadratic behavior for small $J$ is a consequence of the presence of nearly fragmented subspaces, i.e., the slow dynamics controlled by the intra-community process $J$. Our results indicate that for such systems (with $J/t\lesssim0.5$) the $\gamma/J^2$ is constant [see inset of Fig.~\ref{fig10}(b)]. The value of the latter can be obtained from analysis of Eq.~\eqref{dtsecond}. Consider a setup in which the only possible processes are intra-community, given by $J$. Then Eq.~\eqref{dtsecond} can be written as 
\begin{equation}
    \frac{J^2}{4}\frac{1}{Z}\sum_{\alpha \ne \rho} \sum_{i \in C_{\alpha},j \in C_{\rho}} \bra{b_i}H^{\pm}\ket{b_j}=J^2\gamma_{0} \,,
\end{equation}
where $H^{\pm}=\sum_\ell(S^+_\ell S^-_{\ell+1} +S^-_\ell S^+_{\ell+1})$. Consequently, the sums in the above equation just count the number of nonzero elements, $H^{\pm}_{ij}$. The latter can be obtained directly from the  Hamiltonian matrix, without diagonalization. In the inset of Fig.~\ref{fig10}(b) we show that the rate $\gamma$ obtained from fits ${\cal D}(\tau)=1-\gamma\,\tau^2$ perfectly match $\gamma=J^2\gamma_0$, i.e., $\gamma_0=69300/16632/4\simeq 1.04$ for $L=12\,, N_h=2$ and $\gamma_0=10080/3150/4=0.8$ for $L=10\,, N_h=2$. Such behavior breaks down when the division into ${\cal C}_\alpha$ ($\alpha=1,\dots,C$) communities is meaningless, or when the spin-exchange $J$ is large. 

In summary, we have demonstrated that the Hilbert-space structure revealed by spectral graph analysis is fully consistent with the system’s dynamics. Our results indicate a transition between trivial behavior with a single time scale given by both $t$ and $J$, and nearly fragmented behavior in which these scales are separated (i.e., slow dynamics given by $J\ll t$, and fast dynamics given by $t$ for long times). We observe such a transition in both the strength of spin exchange $J$ and in the division into a proper number of communities $C$. The former is presented in Fig.~\ref{fig7}(b) and Fig.~\ref{fig10}(b), indicating nearly fragmented system for $J/t\lesssim 0.5$.  The latter is also shown in Fig.~\ref{fig9}(d), indicating a sudden change when $C>nF$, i.e., when Hilbert space is divided into the "wrong" communities for given $J\ne 0$.

\section{Beyond kinetic fragmentation: Hubbard chain}
\label{sec:hubbard}
In the models considered in the preceding sections, as well as in most models studied in the literature, fragmentation arises from kinetic constraints encoded in the off-diagonal elements of the Hamiltonian, $H_{i\ne j}$, i.e., in the kinetic part of the model. Consequently, the adjacency matrix, constructed exclusively from $H_{i\ne j}$, accurately captures dynamical bottlenecks in fragmented or nearly fragmented systems. However, such models emerge as effective descriptions of more fundamental Hamiltonians, derived from a physical understanding of distinct energy scales. For example, the $t$-$J$ model discussed in previous sections is derived as an effective description of the large-$U$ Hubbard model. This derivation already introduces dynamically disconnected subspaces, each with a fixed number of doubly occupied sites. Then, the $t-J_{z}$ model, based on an additional approximation $J\to0$, introduces further partitioning of these subspaces based on spin configuration. Consequently, the question arises of whether graph partitioning can be applied directly to the fundamental parent model, thereby providing a hierarchy of nearly decoupled subspaces. In the following, we show that it is indeed possible.

To this end, we consider the large-$U$ regime of the Hubbard chain 
\begin{equation}
H=t \left( \sum_{l\sigma}c^\dagger_{l\sigma}c^{\phantom{\dagger}}_{l+1\sigma} + {\rm H.c.} \right) +U \sum_{l} 
n_{l\downarrow} n_{l\downarrow}\;,
\label{eq:hubb}
\end{equation}
where $n_{l\sigma}= c^\dagger_{l\sigma} c _{l\sigma}$ and the Hilbert space includes all states regardless of double occupancy. The first task is to encode the separation of states in the adjacency matrix $A$ arising from different energy scales. Note that there is no unique procedure for incorporating the latter into a graph. Here, we choose a simple condition similar to the perturbative treatment of virtual transitions, i.e., penalizing transitions to high-energy states. Consider matrix elements $H=(H_{ij})$ in basis $\ket{b_i}$. Next, we build the corresponding adjacency matrix $A=(A_{ij})$, where
\begin{equation}
\forall_{i\ne j}\,
A_{ij}=\frac{|H_{ij}|^2}{\mathrm{max}(1,|H_{ii}-H_{jj}|)^2}\,,
\label{eq:hubad}
\end{equation}
and $A_{ii}=0$. Such an adjacency matrix, $A$, can then be used to construct the Laplacian $L$ or modularity $Q$ matrices, which can be finally analyzed in the same way as described in the preceding sections [i.e., Eq.~\eqref{eq:laplacian} and Eq.~\eqref{modularity}]. 

\begin{figure}[!t]
\includegraphics[width=1.0\columnwidth]{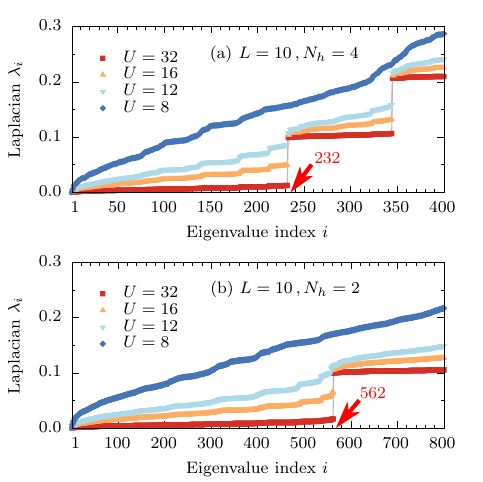}
\caption{Lowest eigenvalues of Laplacian $L$ for the Hubbard model for various strengths of interaction $U/t=32,16,12,8$. Calculated for (a) $L=10$ sites and $N_h=4$ holes, yielding $F=232$ in $U\to\infty$ limit and (b) $L=10\,, N_h=2$ with $F=562$.}
\label{fig11}
\end{figure}
 
 Before performing spectral analysis of graphs corresponding to~\eqref{eq:hubb}, let us first discuss its $U\to\infty$ limit. In the previous sections, we have investigated the $t$-$J_{z}$ limit, i.e., the case with $N_e=L-N_h$ electrons (singlons) and no doubly occupied states (doublons, $N_d=0$). Now, in the Hubbard model, the latter condition is relaxed to the conservation of a number of doublons, i.e., $[H,N_d]=0$. Consequently, one can expect the number of disconnected subspaces to be larger than in the $t$-$J_{z}$ model. To evaluate $F$ in the $U\to0$ limit of the Hubbard model, one can follow three rules derived from first-order processes, i.e., from the direct hopping $t$: 
\begin{itemize}
\item[(1)] a singlon with given spin projection $\sigma$ cannot pass over another singlon $\sigma^\prime$,
\item[(2)] singlons can exchange places with doublons and holons,
\item[(3)] holons and doublons cannot exchange places.
\end{itemize}
Interestingly, these rules encapsulate rich physical phenomena and effective models. For example, from the first rule one derives the $t$-$J_{z}$ model in the sector containing only singlons and holons ($N_d = 0$), which is relevant to high-$T_c$ superconductivity~\cite{Chao1978,Zhang1988}. Systems for which the second rule is relevant - e.g., when the initial state is a mixture of singlons, doublons, and holons - have been shown~\cite{Rigol2004,Rosch2008,HeidrichMeisner2009,Herbrych2017}, including experimentally~\cite{Winkler2006,Xia2015,Scherg2018}, to exhibit a long-lived prethermal state. Finally, from the last rule, in the sector containing only holons and doublons ($N_d = N_e/2$), doublons behave as hard-core bosons, and an effective $XX$-type spin model for the so-called $\eta$-pairs can be derived~\cite{Yang1989,Mark2020,Moudgalya2020}, which is particularly relevant for nonequilibrium photodoping~\cite{Kaneko2019,Murakami2022}. All of these phenomena represent some subspace that is energetically disconnected in the $U\to\infty$ limit (i.e., in the limit where the doublons are conserved), and only a second-order process $\propto t^2/U$ connects them. The number of such decoupled communities can be evaluated as
\begin{equation}
F=\sum_{N_d=0}^{(L-N_h)/2}\binom{L-N_h-2N_d}{(L-N_h-2N_d)/2}\binom{N_h+2N_d}{N_d}\,.
\end{equation}
The first binomial counts the number of $S^z_\mathrm{tot}=0$ spin configurations in which singlons can be placed on $L-N_h-2N_d$ sites with $N_\uparrow=N_\downarrow=(L-N_h-2N_d)/2$ particles of given spin projection. The second binomial counts the number of the holon-dobulon configurations on the remaining $L-(L-N_h-2N_d)=N_h+2N_d$ sites. Note that the above subspaces $C_\alpha$ represent intertwined configurations of spins with doublons and holons. For example, states $\ket{112011}\in C_\alpha$ and $\ket{110211}\notin  C_{\alpha}$ (here $\ket{2}\,, \ket{0}\,, \ket{1}$ represent doublon, holon, and singlon with arbitrary spin $\sigma$ projection, respectively) cannot be connected by the kinetic energy. The transition between these two states would require "melting" of the doublon, $\ket{20}\to\ket{11}\to\ket{02}$, which is a second-order process for large $U$. 

Figure~\ref{fig11} presents the smallest eigenvalues of the Hubbard model Laplacian $\lambda_i$ [built on Eq.~\eqref{eq:hubad}]. For considered systems $L=10\,,N_h=4$ and $L=10\,,N_h=2$ (all in $S^z_\mathrm{tot}=0$ magnetization sector) one expects $232$ and $562$ communities, respectively. Our results indicate that for sufficiently large $U/t\sim 10$, the spectral analysis of the graphs detects all subspaces.

\section{Discussion \& Conclusions}
\label{sec:conclusions}
The graph-based perspective presented here provides a general and flexible tool for exploring Hilbert space fragmentation and its generalization to nearly fragmented systems. This suggests that concepts from network theory can play an important role in understanding complex quantum dynamics. By mapping many-body basis states onto vertices and Hamiltonian matrix elements onto edges, the structure of the Hilbert space is recast as a connectivity problem, allowing one to leverage spectral tools from graph theory. Within this representation, exact fragmentation corresponds to strictly disconnected components, whereas nearly fragmented systems appear as weakly connected communities whose structure is still visible in the graph's spectral properties. A central advantage of this approach is that it does not rely on the prior identification of conserved quantities or model-specific constraints, unlike conventional analyses of fragmentation. Instead, the spectrum of the Laplacian or modularity provides a direct probe of hidden structures in Hilbert space. In particular, small but finite eigenvalues $\lambda$ and/or $\widetilde{\beta}$ signal weak coupling between otherwise isolated sectors, while the associated eigenvectors (such as the Fiedler vector) encode the organization of weakly coupled subspaces. This enables a quantitative distinction between exact fragmentation and regimes where fragmentation is only approximate.

Our results indicate that nearly-fragmented systems form an intermediate regime between fully ergodic dynamics, as described by the ETH, and strictly fragmented models. In this regime, the system exhibits slow dynamics and long-lived memory of initial conditions, and one can expect suppressed transport despite the absence of exact constraints that would prevent thermalization. From a graph-theoretic perspective, this behavior arises from a hierarchical connectivity structure in which strongly connected clusters are linked by sparse or weak edges. Consequently, relaxation proceeds in stages: rapid equilibration within communities is followed by slow exploration of the full Hilbert space, reminiscent of prethermalization phenomena. To connect the spectral properties of graphs with the dynamics of the subspace-projected Loschmidt echo introduced in Chapter \ref{sec:dynamics}, the weights assigned to the graph edges should be chosen to be the squares of the Hamiltonian matrix elements. Another important implication is the robustness of fragmentation-like behavior against weak perturbations. Even when processes are introduced that formally connect all subspaces, the system may remain effectively fragmented on experimentally relevant timescales. This is clearly captured in the spectral structure of the graph, where nearly-degenerate Laplacian and/or modularity eigenvalues and well-defined communities persist despite finite connectivity. Such behavior is particularly relevant to fundamental models of solid-state physics, such as the perturbed $t$-$J$ chain and the Hubbard model studied here, where weak processes couple sectors associated with spin and charge configurations.

The applicability of the present framework extends beyond the specific examples considered in this work. Since the construction relies only on the Hamiltonian matrix in a chosen basis, it can be applied to a wide range of systems, including kinetically constrained models, systems with emergent conservation laws, and models exhibiting constrained dynamics. Moreover, it opens the possibility of employing more advanced tools from network science, such as community detection algorithms or modularity optimization, to further characterize the structure of Hilbert space.

From a phenomenological perspective, the methods introduced here provide a coherent framework for identifying emergent structures in quantum systems across different fragmentation regimes. Although our analysis focused primarily on modularity eigenvectors (i.e., Fig.~\ref{fig5} and Figs.~\ref{fig8}-\ref{fig10}), we find that qualitatively equivalent community structures can also be recovered from Laplacian eigenvectors, provided that the number of clusters used in the subsequent $k$-means procedure is known. The two approaches, however, exhibit distinct practical advantages. Modularity-based spectra are generally more effective in detecting the onset of nearly fragmented communities ($nF>1$), where the residual inter-community couplings remain relevant. In contrast, for fully fragmented systems with $F>1$, the Laplacian approach proves more robust, particularly when the communities differ significantly in size (e.g., as in our Hubbard model investigation, Fig.~\ref{fig11}). This reflects a known limitation of modularity optimization, which tends to overpartition systems containing simultaneously large and small communities \cite{Fortunato2007,Lancichinetti2011}, whereas it performs remarkably well when the relevant structures are of comparable scale. Taken together, these observations suggest that modularity and Laplacian methods should be viewed as complementary probes of emergent organization in quantum systems, with their relative performance determined by the degree of fragmentation and the hierarchy of community sizes.

Finally, several open questions remain. In particular, the dependence of the graph representation on the chosen basis deserves further investigation, as physically meaningful fragmentation depends on this choice. Additionally, scaling the method to larger systems will require approximate or coarse-grained graph constructions, which could provide an avenue for studying fragmentation in regimes beyond exact diagonalization. In this context, it is worth noting that the spectral diagnostics employed here rely primarily on extremal eigenvalues and their associated eigenvectors, making them well-suited for iterative methods such as Lanczos or Chebyshev-based algorithms. These approaches offer a promising route to accessing the relevant spectral features in much larger Hilbert spaces without requiring full diagonalization. Future work may also extend to dynamical observables, investigate the stability of nearly-fragmented structures under stronger perturbations, and explore connections to other forms of slow dynamics and nonthermal behavior in quantum many-body systems.

An interesting direction for future work is to explore complementary formulations of fragmentation in terms of binary $(0,1)$ variables, such as those arising in graph partitioning or modularity maximization. These problems are known to be NP-hard, placing them in a different complexity class from the spectral analysis considered here, which can be solved in polynomial time. In this context, it is natural to consider mappings to quadratic unconstrained binary optimization (QUBO) problems, which provide a unifying framework for discrete optimization and can be addressed using both classical heuristics and emerging quantum optimization methods. Such approaches could, in principle, offer new ways to identify optimal partitions of Hilbert space \cite{Ushijima2017,Cheng2025} and to sharpen the characterization of nearly-fragmented regimes. Exploring the interplay between these discrete formulations and the continuous spectral perspective developed here may therefore open a promising avenue for future research.

\begin{acknowledgments}
We thank F.~Heidrich-Meisner, P.~Łydżba, and J.~Bon\v{c}a for fruitful discussions. J.H. acknowledges support by the National Science Centre (NCN), Poland, via project 2023/50/E/ST3/00033. The calculations have been carried out using resources provided by the Wroclaw Centre for Networking and Supercomputing (\url{http://wcss.pl}). The data that support the findings of this article are openly available \cite{opendata}.
\end{acknowledgments}

\appendix

\section{Spectral partitioning of graphs}
\label{sec:ap1}
In case of a strictly fragmented system, the partition aims to divide the Hilbert space into {\em two} subspaces, ${\mathcal H}={\mathcal H}_1 \oplus {\mathcal H}_2$ that are not connected by any nonzero matrix element, so that 
\begin{equation}
\lambda= \frac{4}{{\rm dim}({\cal H})}\sum_{\ket{b_i} \in {\mathcal H}_1,\ket{b_j}\in {\mathcal H}_2} A_{ij}=0\;, 
\label{s1.rdef}
\end{equation} 
where the constant factor $4/{\rm dim}({\cal H}) $ is introduced only for convenience. Each partitioning can be represented by a vector
\mbox{$\mathbf{v}^T=(v_1,v_2,v_3,\dots,v_{\mathrm{dim}({\cal H})})$} with components $v_i=1$ for $\ket{b_i} \in {\mathcal H}_1$ and $v_i=-1$ for $\ket{b_i}\in {\mathcal H}_2$. The key step \cite{Fiedler1973} is to note that \mbox{$(1-v_i v_j)/2=0$} if $\ket{b_i} $ and $\ket{b_j}$ belong to the same subspace and \mbox{$(1-v_i v_j)/2=1$} otherwise. Using these identities, one may perform the summation in Eq.~\eqref{s1.rdef} over all basis states
\begin{eqnarray}
\lambda &= &\frac{2}{{\rm dim}({\cal H})}\sum_{i,j} \frac{1-v_i v_j}{2} A_{ij}\nonumber \\
&= &\frac{1}{{\rm dim}({\cal H})}\sum_{i,j}v_i(D_i \delta_{ij} -A_{ij})v_j \nonumber \\
&=&\frac{\mathbf{v}^T L \mathbf{v}}{\mathbf{v}^T \mathbf{v}}=0\;.
\label{s1.rdef2}
\end{eqnarray} 
Here, we introduced the Laplacian matrix, $L=D-A$, with a diagonal part,
$D=\mathrm{diag}(d_1,d_2,d_3,\dots,d_{\mathrm{dim}({\cal H})})$, determined by the degree of vertices, $d_i=\sum_j A_{ij}$. Eq.~\eqref{s1.rdef2} shows that partitioning is represented by a vector, $\mathbf{v}$, that belongs to the degenerate subspace associated with the zero eigenvalue of the Laplacian matrix. 

Consider now a case when the Hilbert space consists of $F$ disconnected subspaces, ${\cal H}_{\alpha}$, $\alpha=1,\dots,F$. Then all the basis states in each ${\cal H}_{\alpha}$ can be labeled by either $v_i=1$ or $v_i=-1$ (with the same choice for all states in a given subspace). In this way, one generates $2^F/2$ distinct bipartitions, and ${\cal R}=0$ for each bipartition. Consequently, all such $\bf v$ belong to the degenerate subspace, corresponding to a vanishing eigenvalue of the Laplacian. However, the vectors $\bf{v}$ are not orthogonal, so that the dimension of this degenerate subspace is smaller than $2^F/2$. To obtain its dimension, we note that all such eigenvectors $\mathbf{v}$ can be represented as linear combinations of $F$ {\it orthonormal} vectors $\mathbf{u}^{\alpha}$. Here, $(\mathbf{u}^{\alpha})^T=(u^{\alpha}_1,u^{\alpha}_2,u^{\alpha}_3,\dots,u^{\alpha}_{\mathrm{dim}({\cal H})})$, where $u^{\alpha}_i=1$ for $\ket{b_i} \in {\cal H}_{\alpha}$ and $u^{\alpha}_i=0$ for $\ket{b_i}\notin {\cal H}_{\alpha}$. Consequently, one obtains an $F$-dimensional degenerate subspace. In other words, the number of disconnected subspaces, ${\cal H}_{\alpha}$, of the fragmented model equals the degeneracy of the zero eigenvalue of the Laplacian matrix. 

An analogous spectral analysis can be performed for the modularity matrix defined in Eq.~\eqref{eq:mod} of the main text. Here, one looks for a bipartition, ${\mathcal H}={\mathcal H}_1 \oplus {\mathcal H}_2$, that maximizes
\begin{eqnarray}
\beta&=& \frac{2}{{\rm dim}({\cal H})}\left[ \sum_{i,j \in {\cal H}_1 }Q_{ij}+\sum_{i,j \in {\cal H}_2}Q_{ij}
\right] \nonumber \\
&=& \frac{2}{{\rm dim}({\cal H})} \sum_{i,j} Q_{ij} \frac{1+v_i v_j}{2} \nonumber \\
&=& \frac{\mathbf{v}^T Q \mathbf{v}}{\mathbf{v}^T \mathbf{v}}
+ \frac{1}{{\rm dim}({\cal H})} \sum_{i,j} Q_{ij} \label{s1.mod2}\;,
\end{eqnarray}
where we assumed that the graph partition is encoded in $v_i$ in the same way as described below Eq.~\eqref{s1.rdef} for the Laplacian matrix, i.e., $v_i=1$ for $\ket{b_i} \in {\cal H}_1$ and $v_1=-1$ for $\ket{b_i} \in {\cal H}_2$. The first term in Eq.~\eqref{s1.mod2} shows that the partition that maximizes $\beta$ is encoded in the eigenvector corresponding to the largest eigenvalue, $\beta_{\rm max}$, of $Q$. One may also easily check that the second term in Eq.~\eqref{s1.mod2} vanishes. For the same reason, the trivial solution of Eq.~\eqref{s1.rdef2}, $\mathbf{v}_1^T=(1,1,1,\dots)$, is eliminated from the approach based on the modularity matrix since the corresponding eigenvalue equals $\beta=1/{\rm dim}({\cal H}) \sum_{i,j} Q_{i,j}=0 < \beta_{\rm max}$. Consequently, for a fragmented or nearly fragmented system with $F$ subspaces, one looks for the $F$ smallest eigenvalues and eigenvectors of the Laplacian matrix or $F-1$ largest eigenvalues and eigenvectors of the modularity matrix. 

\section{Modularity ${\cal R}$}
\label{sec:ap2}
In this section, we estimate the changes in the modularity eigenvalues arising from a fragmentation-breaking perturbation. Let us assume that $\beta$ is the largest solution of Eq.~\eqref{s1.mod2} for a perfectly fragmented system with adjacency matrix $A$, modularity matrix $Q$, and the corresponding partitioning, ${\cal H}={\cal H}_1 \oplus {\cal H}_2$, is encoded in vector $\bf v$. The same quantities for the perturbed system will be denoted by $\beta'$, $A'$,
$Q'$, and 
$\bf v'$, respectively. We introduce a notation 
\begin{equation}
a_{\alpha,\beta} = \sum_{\ket{b_i} \in {\mathcal H}_{\alpha},\ket{b_j}\in {\mathcal H}_{\beta}} A_{ij}\,, 
\end{equation}
for which we find
\begin{eqnarray}
\beta&=&\frac{1}{{\rm dim}({\cal H})} \left(
a_{11}+a_{22}-2a_{12}
-\frac{(a_{11}-a_{22})^2}{a_{11}+a_{22}+2a_{12}}
\right)\nonumber \\
&=& \frac{4}{{\rm dim}({\cal H})} \; \;
\frac{a_{11}a_{22}-a^2_{12}}{a_{11}+a_{22}+2a_{12}}\,. \label{s1.mod3}
\end{eqnarray}
Strict fragmentation means that $a_{12}=0$. Next, we consider the simplest case in which the perturbation introduces a nonvanishing $a_{12} > 0$ without modifying $a_{11}$ or $a_{22}$. In other words, such a perturbation couples subspaces without changing the matrix elements within the subspaces.  In this case, one finds the following inequalities
\begin{eqnarray}
\beta' & = & \frac{\mathbf{v'}^T Q' \mathbf{v'}}{\mathbf{v'}^T \mathbf{v'}} \ge \frac{\mathbf{v}^T Q' \mathbf{v}}{\mathbf{v}^T \mathbf{v}}\,,\nonumber \\
\beta' & \ge & \frac{1}{{\rm dim}({\cal H})} \left(
a_{11}+a_{22}
-\frac{(a_{11}-a_{22})^2}{a_{11}+a_{22}}
\right)-\frac{2a_{12}}{{\rm dim}({\cal H})}\,,\nonumber \\
\beta' & \ge & \beta - {\cal R}/2\,, \label{s1.mod4}
\end{eqnarray}
where ${\cal R}$ was defined in Eq.~\eqref{eq:l1} in the context of the Laplacian matrix.  Eq.~\eqref{s1.mod3} shows that weak coupling between the subspaces, $a_{12}\ne 0$, tends to reduce $\beta$, while Eq.~\eqref{s1.mod4} shows that this reduction should be of the same order of magnitude as changes introduced to the spectra of the Laplacian matrix. See Eq.~\eqref{eq:l1}.

\section{Short time expansion}
\label{sec:ap3}
We will now show that the dynamics of the subspace projected Loschmidt echo [Eq.~\eqref{do2} of the main text] is possible only when the subspaces are mixed. Consider Eq.~\eqref{do1} and a state $\ket{b_j}\in {\cal C}_{\alpha}$
\begin{eqnarray}
{\cal D}_{\ket{b_j}}(\tau)&=&\sum_{\ket{a_j}\in {\cal C}_{\alpha}}
|\bra{a_j} U(\tau)\ket{b_j}|^2\,,\nonumber\\
&=&\sum_{\ket{a_j}\in {\cal C}_{\alpha}}
\bra{b_j} U^\dagger(\tau)\ket{a_j} \bra{a_j} U(\tau)\ket{b_j}\nonumber\\
&=&\bra{b_j} U^\dagger(\tau)\,\Pi_\alpha\, U(\tau)\ket{b_j}\,,
\end{eqnarray}
where, $U(\tau)=\exp(-iH\tau)$ , in the last equation we introduced the projector into ${\cal C}_{\alpha}$ subspace
\begin{equation}
\Pi_\alpha=\sum_{\ket{a_j}\in {\cal C}_{\alpha}}\ket{a_j} \bra{a_j}=\Pi_\alpha^2\,.
\end{equation}
Now, Eq.~\eqref{do2} becomes
\begin{eqnarray}
{\cal D}(\tau)&=&\frac{1}{Z} \sum_{\alpha} \sum_{\ket{b_j}\in {\cal C}_{\alpha}} {\cal D}_{\ket{b_j}}(\tau) \,,\nonumber\\
&=&\frac{1}{Z}\sum_{\alpha}\sum_{\ket{b_j}\in {\cal C}_\alpha} \bra{b_j} U^\dagger(\tau)\,\Pi_\alpha U(\tau)\ket{b_j}\nonumber\\
&=&\frac{1}{Z}\sum_{\alpha}\sum_{\ket{b_j}}\bra{b_j}\Pi_\alpha U^\dagger(\tau)\,\Pi_\alpha U(\tau)\ket{b_j}\nonumber\\
&=&\frac{1}{Z}\sum_{\alpha}\Tr\left[
\Pi_\alpha U^\dagger(\tau)\,\Pi_\alpha U(\tau)
\right]\,.
\end{eqnarray}
Here we have used $Z=\sum_\alpha\mathrm{dim}(C_\alpha)$ and 
\begin{equation}
\sum_{\ket{b_j}\in {\cal C}_\alpha} \ket{b_j}=\sum_{\ket{b_j}}\Pi_\alpha\ket{b_j}\,.
\end{equation}
Expanding the time evolution operator
\begin{equation}
U(\tau)=\mathds{1}-i\tau H-\tau^2 H/2+{\cal O}(\tau^3)
\end{equation}
one gets [up to ${\cal O}(\tau^3)$ terms]
\begin{equation}
{\cal D}(\tau)=\frac{1}{Z}\sum_{\alpha}\Tr\left[
\Pi_\alpha-\tau^2\Pi_\alpha H^2 + \tau^2\Pi_\alpha H\,\Pi_\alpha H
\right]\,.
\end{equation}
Inserting $\mathds{1}=\Pi_\alpha+(\mathds{1}-\Pi_\alpha)$ into $H^2=H\mathds{1}H$ yields
\begin{eqnarray}
{\cal D}(\tau)=\frac{1}{Z}\sum_{\alpha}\Tr\left[
\Pi_\alpha - \tau^2\,\Pi_\alpha H(\mathds{1}-\Pi_\alpha)H
\right]\,. \label{dt1}
\end{eqnarray}
Next we split the Hamiltonian into two parts, \mbox{$H=(\sum_{\alpha} \Pi_{\alpha})H(\sum_{\rho} \Pi_{\rho})=H_0+H'$} which, respectively, preserve and break the fragmentation 
\begin{eqnarray}
H_0&=&\sum_{\alpha} \Pi_\alpha H   \Pi_\alpha \nonumber \\
H'&=&\sum_{\alpha \ne \rho} \Pi_\alpha H   \Pi_\rho\; . \nonumber \\
\end{eqnarray}
Using the identities, $[\Pi_{\alpha},H_0]=0$, 
$\Pi_{\alpha}H'\Pi_{\alpha}=0$, and 
$\Pi_{\alpha}(1-\Pi_{\alpha})=0$ one can simplify the short-time expansion from Eq.\eqref{dt1},
\begin{equation}
{\cal D}(\tau)=1-\tau^2\frac{1}{Z} \Tr[(H')^2]   
\end{equation}
where
\begin{equation}
 \Tr[(H')^2]=\sum_{\alpha \ne \rho} \sum_{i \in C_{\alpha},j \in C_{\rho}} |\bra{b_i}H \ket{b_j}|^2 \;.
\end{equation}
Note that the short-time dynamics of ${\cal D}(\tau)$ is given by the Hilbert-Schmidt norm of the perturbation that breaks the strict fragmentation. The very same quantity also appears in the estimates of the small (large) eigenvalues of the Laplacian (modularity) matrix. It explains why the spectral analysis based on the seemingly unrelated graph partitioning is reflected in the actual quantum dynamics.

\bibliography{graphs.bib}
\end{document}